\documentclass[twocolumn,showpacs,prb]{revtex4-1}

\usepackage{graphicx}
\usepackage{color}\usepackage{enumerate}
\usepackage{amsmath}\usepackage{amssymb}\usepackage{stmaryrd}
\usepackage{multirow}
\usepackage{float}
\usepackage{esvect}
\usepackage{url}
\usepackage{hyperref}
\usepackage[normalem]{ulem}
\usepackage[mathscr]{eucal}
\begin{document}

\title{Anyonic Symmetries and Topological Defects in Abelian Topological Phases: an application to the $ADE$ Classification}

\author{Mayukh Nilay Khan}
\author{Jeffrey C.Y. Teo}%\email{cteo@illinois.edu}
\author{Taylor L. Hughes}
\affiliation{Department of Physics, Institute for Condensed Matter Theory, University of Illinois at Urbana-Champaign, IL 61801, USA}
\newcommand\alignedbox[2]{
  % #1 = before alignment
  % #2 = after alignment
  &
  \begingroup
  \settowidth\dlf{$\displaystyle #1$}
  \addtolength\dlf{\fboxsep+\fboxrule}
  \hspace{-\dlf}
  \boxed{#1 #2}
  \endgroup
}

%%%%%%%%%%%Outline%%%%%%%%%%%%%
%anyon relabeling symmetry of ADE topological phases; Outer automorphism of Lie algebra, symmetry of the Dynkin diagram;
%gapped interface phases, distinct phases labeled by anyonic symmetry
%gapped edge phases for fractional quantum spin Hall
%twist defects: non-Abelian <-> multi-channel fusion, fraction Ising like;
%triality symmetry of so(8) state, fractional quantum Hall, lattice model, surface of bosonic symmetry protected phase; non-Abelian symmetry group -> non-commutative defect fusion

%%%%%A more precise but less general abstract%%%%%%%
%We study symmetries and defects of a wide class of Abelian topological phases whose edges exhibit Lie algebraic conformal structures. We formulate the group structures of anyon relabeling symmetries of all Abelian topological phases. This in particular applies to phases under the Cartan $ADE$ classification of simply-laced Lie algebras, and their anyonic symmetries coincide with reflections or rotations of Dynkin diagrams also known as outer automorphisms. Topological defects are constructed as domain walls between gapped interface phases and are classified according to anyonic symmetries and species of bound quasiparticles. We derive the defect fusion properties in all $ADE$ topological states, and show their non-Abelian fractional Ising-like fusion characteristics.

%%%%%A more general but less precise abstract%%%%%%%%
\begin{abstract}
We study symmetries and defects of a wide class of two dimensional Abelian
topological phases characterized by Lie algebras. We formulate the symmetry
group of all Abelian topological field theories. The symmetries relabel
quasiparticles (or anyons) but leave exchange and braiding statistics unchanged.
Within the class of $ADE$ phases in particular, these anyonic symmetries have a
natural origin from the Lie algebra. We classify one dimensional gapped phases
along the interface between identical topological states according to symmetries.
This classification also applies to gapped edges of a wide range of fractional
quantum spin Hall(QSH) states. We show that the edge states of the $ADE$ QSH
systems can be gapped even in the presence of time reversal and
charge conservation symmetry. We distinguish topological point defects
according to anyonic symmetries and bound quasiparticles. Although in an Abelian
system, they surprisingly exhibit non-Abelian fractional Majorana-like
characteristics from their fusion behavior.
\end{abstract}

\maketitle
\section{Introduction and Motivation}
Topologically ordered phases with Abelian anyons are usually considered to be the simplest examples of topological order (TO), however recent exciting work has shown that the theory is still far from complete. Two notable developments related to our current work are: (i) the generation of semi-classical defects in Abelian topological phases that exhibit similar features to non-Abelian quasiparticles\cite{Kitaev06, barkeshli2010, Bombin, Bombin11, KitaevKong12, kong2012A,YouWen, YouJianWen, BarkeshliQi, BarkeshliQi13, BarkeshliJianQi, MesarosKimRan13, TeoRoyXiao13long, teo2013braiding} and (ii) the  bulk-boundary correspondence for topological phases with and without symmetry protection, and the resulting stability of the gapless edge theories~\cite{Haldane95, LevinGu12, Levin13, BarkeshliJianQi13, BarkeshliJianQi13long, cano2013bulk, LuLee13}.%\cite{levin,chetan,yuanming,etc}.
 
Some aspects of these two lines of research can be unified by applying the concept of \emph{anyonic} symmetry (AS). TO phases  support an additional AS structure if the quasiparticle (QP) fusion and braiding are invariant under a set of anyon relabeling operations. This is a common feature in many topological states, such the Kitaev toric code~\cite{Kitaev97} which has an electromagnetic-duality, and the  Abelian $(mmn)$-fractional quantum Hall (FQH) states~\cite{Wentopologicalorder90, WenchiralLL90, WenZee92, Wenbook, Fradkinbook, BoebingerJiangPfeifferWest90, SuenJoSantosEngelHwangShayegan91, EisensteinBoebingerPfeifferWestHe92} which have a bi-layer symmetry. An element of the AS group might, for example, switch a particular anyon-type between the two layers in bi-layer FQH states. The AS is not necessarily a symmetry of the quantum Hamiltonian, but rather a symmetry of the anyon content. 
%For example there is no general symmetry operation that interchanges
%vertices and plaquettes of the Kitaev toric code on an irregular simplicial network. 
For example, an AS could permute QP excitations with different energies. 
%In fact the
%details of a Hamiltonian in a topological field theory is irrelevant as long as it provides a finite excitation gap. 
In general, a ground state in a closed system will not be invariant under an AS operation, and therefore the symmetry can be regarded as being {\em weakly} broken~\cite{Kitaev06,wangLevin2013weak}. However, unlike a classical symmetry-broken phase, the AS may not be associated with a physical quantity, and cannot be measured by a finite vacuum expectation value of any local observable. 

In this work we construct a class of Abelian bosonic FQH states associated with elements of the $ADE$ Cartan classification of Lie-algebras and show that they have AS. For these systems the AS can be used to create non-Abelian \emph{twist defects} and topologically distinct gapped edge, or interface phases. In addition we can apply our result to predict when edge theories of certain time-reversal invariant fractional quantum spin Hall states (FQSH) made from time-reversed copies of the $ADE$ FQH states can be gapped without breaking symmetries.
One remarkable result we find is an exact mapping between the well-known \emph{triality} symmetry of the Lie algebra $so(8)$ and the AS of the associated topological state. In fact, we prove that the AS for these theories are exactly the symmetries of the Dynkin diagrams that represent the ADE Lie algebras. This is not only applicable to a 2D FQH state that carries an $so(8)$ edge algebra, but also the spin liquid surface state of a three dimensional bosonic symmetry protected phase~\cite{VishwanathSenthil12, WangSenthil13, BurnellChenFidkowskiVishwanath13, WangPotterSenthil13}. 

%\begin{itemize}
%\item examples of anyonic symmetry toric code\cite{Kitaev97}, Bombin color code\cite{BombinColorCode}, bilayer FQH, topological nematic states\cite{BarkeshliQi}.
%\item topologically ordered gapped surface of 3D TI - no spontaneously $U(1)$ and TR symmetry breaking but impossible to realize TR as an (antiunitary) anyonic symmetry compatible with
%charge conservation
%\item examples of twist defects, dislocation in toric code\cite{KitaevKong12,Bombin}, genons\cite{BarkeshliJianQi}, disclinations in color code\cite{TeoRoyXiao13long}, bilayer quantum Hall
%states\cite{BarkeshliQi,teo2013braiding},$\mathbb{Z}_k$ rotor model. \cite{YouWen}
%\end{itemize}

%%%%%%%%%%%%%%%% Introduction %%%%%%%%%%%%%%%%
To begin, we need to introduce the well-known $K$-matrix formalism for Abelian
TO states. An Abelian FQH state is described by an effective Chern-Simons
topological field theory $\mathcal{L}=\frac{1}{4\pi}K_{IJ}\alpha_I\wedge
d\alpha_J$ in $2+1$ dimensions, where $\alpha_I$ is an $r$-component set of $U(1)$
gauge fields. The topological state is characterized by the symmetric,
integral-valued $K$-matrix~\cite{WenZee92}. QP excitations of the theory are
labeled as $r$-component vectors (${\bf{a,b}}\ldots$) in an integer (anyon)
lattice $\Gamma^\ast=\mathbb{Z}^r.$ Vector addition corresponds to QP fusion
$\psi^{\bf a}\times\psi^{\bf b}=\psi^{{\bf a}+{\bf b}}$.

The spin $h_{\bf{a}}$ of a QP $\psi^{\bf a}$ is given by $\frac{1}{2}\mathbf{a}^tK^{-1}\mathbf{a}$.
The topological spin (or exchange
statistics) of a QP $\psi^{\bf a}$ is given by $\theta_{\bf a}=e^{\pi i{\bf
a}^TK^{-1}{\bf a}}$, and encircling a QP $\psi^{\bf a}$ once around another QP
$\psi^{\bf b}$ gives the braiding phase $\mathcal{D} S_{{\bf a}{\bf b}}=e^{2\pi
i{\bf a}^TK^{-1}{\bf b}}$, for $\mathcal{D}=\sqrt{|\det(K)|}\geq1$.
The topological spin of the quasiparticles is often stated in terms of the $T$
matrix,
$T_{\mathbf{a}\mathbf{b}}=e^{2\pi i h_{\mathbf{a}}}=\delta_{\mathbf{a},\mathbf{b}}\theta_{\mathbf{a}}$. 
We note that for completely chiral theories, (using bulk boundary correspondence) 
the spin $h_{\bf{a}}$ is the same as the scaling dimension of the primary 
fields(corresponding to the quasiparticles) of the (1+1-d) conformatl field theory (CFT) on the edge.

The QPs that occupy the sublattice $\Gamma=K\mathbb{Z}^r\subseteq\Gamma^\ast$ are called \emph{local} and only contribute trivial braiding phases with all other QPs. Intuitively they are the fundamental building blocks that are ``fractionalized" to form the topological state; we will enforce that all local particles be bosonic by requiring the diagonal entries of $K$ be even. Topological information encoded in the nonlocal braiding and exchange statistics 
of fractionalized QPs is left invariant upon the addition of local particles $\in\Gamma$.  We can remove this redundancy by labeling distinct
QPs with elements of the anyon quotient lattice $\mathcal{A}=\Gamma^\ast/\Gamma=\mathbb{Z}^r/K\mathbb{Z}^r$. 

QPs are electromagnetically charged in the presence of the additional coupling 
term $\frac{e^\ast}{2\pi}t_IA\wedge d\alpha_I$ where $A$ is the external electromagnetic gauge field and $e^\ast$ is the unit charge of the fundamental local boson. We will assume a symmetric coupling ${\bf t}=(t_I)=(1,\ldots,1)$ which, for example,  is the natural choice in multi-layer systems. The charge of a QP $\psi^{\bf a}$ is $q_{\bf a}=e^\ast{\bf t}^TK^{-1}{\bf a}$.
At zero temperature, the ground state is a Bose-Einstein condensate of local bosons with 
 broken $U(1)$ symmetry/number conservation.
 Physically, the boson condensate could describe an {\em anyonic superconductor}~\cite{Wilczekbook, Preskilllecturenotes} where local particles are Cooper pairs of electrons or perhaps a strongly correlated cold atomic system.
 QPs that differ by local bosons are indistinguishable and interchangeable up to the boson condensate vacuum. Thus, due to the boson condensate, the QP charges $q_{\bf a}$ are only defined modulo integral units of $e^\ast$ at zero temperature. 
 This motivates an intuitive way to think about the anyon quotient lattice as $\mathbb{Z}^r/K\mathbb{Z}^r$, i.e., the anyons are only defined modulo the local bosons that make up the lattice $K\mathbb{Z}^r.$

%%%%%%%%%%%%%%%% ADE classification %%%%%%%%%%%%%%%%
We are interested in Abelian topological states which carry chiral Kac-Moody (KM) current algebras at level 1 along their edges. These include a range of FQH states under the Cartan $ADE$ classification of simply-laced Lie algebras~\cite{mathieu1997conformal, FrohlichThiran94, FrohlichStuderThiran97}. 
The set of $A_r$ and $D_r$ form infinite sequences while there are only three exceptional $E_{r=6,7,8}$. 
In this article, we consider Abelian topological states where the $K$-matrix is given by the Cartan matrix of a corresponding (simply-laced) Lie algebra (which has rank $r$). We will henceforth refer to these models as
$ADE$ states since the Lie algebras with symmetric Cartan matrices that are suitable to form K-matrices are the $A_n, D_n$ and $E_n$ series from the Cartan classification of Lie algebras. This construction
ensures the presence of a Kac-Moody algebra corresponding to the same Lie algebra at level $1$ at the edge of the system, and that the modular $S$ matrix of the bulk topological
state is the same as that of the corresponding affine Lie Algebra at level $1$.\cite{mathieu1997conformal} Strictly speaking, the presence of a sector whose $K$ (sub-)matrix is identical to
the Cartan matrix of the Lie algebra (i.e. we can identify a sub-matrix of the full $K$-matrix which is the Cartan matrix of the Lie algebra) is enough to define a set of currents which obey  Kac-Moody algebra at level $1$ at its edge. 
\cite{mathieu1997conformal, FrohlichThiran94, FrohlichStuderThiran97,cappelli1995stable}. Examples of such $K$-matrices are given 
in Eqs. \eqref{eq:Kmat1},\eqref{eq:Kmat2}. However, we will not explicitly deal with such states in this manuscript.

In the remainder of this section we will motivate the reasons for studying the $ADE$ fractional quantum Hall states.  In fact, these fractional quantum Hall states are relevant to several important lines of research. 
The importance of the states in the $A$ series for the study of FQH hierarchy states
 has already been pointed out in a series of papers as early as Ref. \onlinecite{Readhierarchy} and 
the structure was further clarified by \cite{FrohlichZee,FrohlichThiran94,FrohlichStuderThiran97}.
 This is particularly relevant for the hierarchy states at filling fractions
 $\frac{m}{mp\pm1}$ with symmetry $u(1)\times su(m)_1$ \cite{cappelli1995stable}, where $p$ is 
 an even positive integer. The factor of $su(m)_1$ is exactly the symmetry of the $A$ series of Lie algebras. 
 
As an example, the second hierarchy state at filling fraction $\nu=\frac{2}{2\times 2+1}=\frac{2}{5}$ has 
  \begin{align}
  K=\left(
\begin{array}{cc}
 3 & -1 \\
 -1 & 2 \\
\end{array}
\right)
\label{eq:Kmat1}
\end{align}
with charge vector $t=\begin{pmatrix}{}1\\0\end{pmatrix}$
and  symmetry $u(1)\times su(2)_1$.
The third hierarchy state at filling fraction $\nu=\frac{3}{3\times 2-1}=\frac{3}{5}$
has symmetry $u(1)\times su(3)_1$, K matrix
\begin{align}
\left(
\begin{array}{ccc}
 1 & -1 & 0 \\
 -1 & -2 & 1 \\
 0 & 1 & -2 \\
\end{array}
\right),
\label{eq:Kmat2}
\end{align}
and  a charge vector $t=\begin{pmatrix}{}1\\0\\0\end{pmatrix}.$
 The general structure in these cases  follows this basic pattern. 
 We also note that Laughlin states are \emph{stably} equivalent to $su(n)_1$ states in the $A$ series as was explicitly noted for the $\frac{1}{3}$ state in Ref. \onlinecite{cano2013bulk}.
 In our work we will not focus on the hierarchy states because  these have  an extra fermionic $u(1)$ (charge) sector which complicates the anyonic symmetry analysis to follow.  We instead present the bosonic case first and then discuss the strategy to solve  the fermionic problem in future work. 
 
Beyond their relevance for hierarchy states, the $ADE$ states have also been featured in the recent discussion of topologically ordered and symmetry protected topological states. For example, the $E_8$ state has  been in focus because it is a bosonic short range entangled phase with no topological order\cite{LuVishwanathE8}. The $so(8)$ state, which lies in the $D$ series,  exists on the surface of a 3D
 bulk Symmetry protected topological(SPT) phase protected by time reversal symmetry\cite{BurnellChenFidkowskiVishwanath13,WangSenthil13,WangPotterSenthil13}.  As we will soon show,  the $so(8)$ state has very
 special properties in the context of anyonic symmetry that could be probed if such a bulk 3D topological phase were discovered. Additionally, as mentioned further below, the $D$ series forms half of the sixteen-fold classification scheme first predicted by Kitaev for 2D topologically ordered states\cite{Kitaev06}, and further considered in recent work on interacting 2D\cite{GuLevin14,QiZ8} and 3D topological phases\cite{LukaszChenVishwanath,metlitski2014interaction,
 WangSenthil14}. 
 
 From a mindset of pure convenience, the $ADE$ states also have the advantage that their anyonic symmetries can be fully classified using powerful results from the mathematics of Lie algebras as we will see, whereas  identifying the exact anyonic symmetry group is still an open problem for states with generic $K$-matrices/anyon lattices. With these motivations in mind we will now proceed to the discussion of anyonic symmetry and its consequences.

\section{Anyonic Symmetries of the $ADE$-series}

\begin{figure}[htbp]
	\begin{center}
		\includegraphics[width=0.4\textwidth]{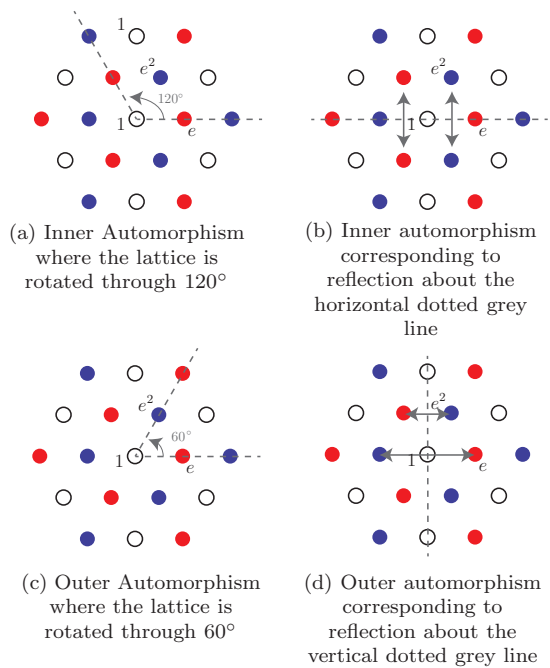}
	\end{center}
	\caption{Anyon lattice of $su(3)$ with inner and outer automorphisms. The white, red and 
blue circles refer to the distinct quasiparticles $1$, $e$ and $e^2$ respectively 
in the anyon quotient lattice 
$\mathbb{Z}^2/K\mathbb{Z}^2$. 
(a) and (b) are examples of inner automorphisms where anyon 
labels (colors of the circles) are preserved whereas (c) and (d) 
the outer automorphisms exchange $e$ and $e^2$ (blue and red circles)}
\label{fig:su3lattice}
\end{figure}

Throughout this article we will use two explicit examples of ADE states to illustrate our results: $A_2=su(3)$ and
$D_4=so(8)$  which are described by \begin{align}K_{su(3)}=\left[\begin{array}{*{20}c}2&-1\\-1&2\end{array}\right],\quad
K_{so(8)}=\left[\begin{array}{*{20}c}2&-1&-1&-1\\-1&2&0&0\\-1&0&2&0\\-1&0&0&2\end{array}\right].\label{KmatricesA2D4}\end{align}
%K_{so(8)}=\left[\begin{array}{*{20}c}2&-1&0&0\\-1&2&-1&-1\\0&-1&2&0\\0&-1&0&2\end{array}\right]
The $su(3)$ state has 3 QPs: $1=(0,0)$, $e=(-1,1)$ and $e^2=(1,-1)$, which form the anyon quotient lattice
$\mathcal{A}_{su(3)}=\mathbb{Z}_3$ with fusion $e\times e=e^2$ and $e\times
e^2=1$ up to local bosons. The QPs have neutral electric charge, but have non-trivial spin 
$\theta_e=\theta_{e^{2}}=e^{2\pi i/3}$ and braiding phases
$\sqrt{3}S_{ee}=\sqrt{3}S_{e^2e^2}=e^{-2\pi i/3}$ and $\sqrt{3}S_{ee^2}=e^{2\pi i/3}$.
The anyon quotient lattice of $su(3)$ is two dimensional and can be easily illustrated in 2D. 
It is shown in Fig. \ref{fig:su3lattice} where the white, red and blue circles refer to the quasiparticles
$1$, $e$ and $e^2$.  The inner product between vectors $\mathbf{a},\mathbf{b}$ of the 
lattice is given by $\mathbf{a}^TK^{-1}_{su(3)}\mathbf{b}$. Thus, the inner product between vectors $(1,0)$($\equiv e$) and $(0,1)$($\equiv e^2$) is 
$\langle(1,0),(0,1)\rangle=\frac{1}{2}$ prompting the representation as a triangular lattice.

The $so(8)$ state has 4 QPs: $1=(0,0,0,0)$, $e=(0,-1,0,1)$, $m=(0,-1,1,0)$ and $\psi=(0,0,-1,1)$ forming the anyon quotient lattice
$\mathcal{A}_{so(8)}=\mathbb{Z}_2\times\mathbb{Z}_2$ with fusion rules $e^2=m^2=1$ and $\psi=e\times m$ up to local bosons. They have neutral electric charge, carry fermionic spin
$\theta_e=\theta_m=\theta_\psi=-1,$ and braiding phases $2S_{ee}=2S_{mm}=2S_{\psi\psi}=1,$ and $2S_{em}=2S_{m\psi}=2S_{\psi e}=-1$. Thus, the three non-trivial anyons are fermions.

The anyon labels and fusion rules for general $ADE$ states are listed in Table~\ref{tab:quasiparticles} and Table \ref{tab:Quasiparticle_labels}. 
In the Lie algebra language the lattice $\Gamma=K\mathbb{Z}^r$ of local bosons is the \emph{root} lattice, while the anyon lattice $\Gamma^\ast=\mathbb{Z}^r$ is called the \emph{weight} lattice~\cite{cano2013bulk} and is dual to $\Gamma$ under the bilinear product $\langle{\bf{a}},{\bf{b}}\rangle={\bf{a}}^TK^{-1}{\bf{b}}$.
%{\bf{TLH: What does dual under $K^{-1}$ mean precisely?}} 
We have omitted the $E_8$ state with trivial topological order
($\mathcal{D}=1$)~\cite{LuVishwanathE8}. We note in passing that there is an
eightfold periodicity in the $D_r$ series with rank $r\geq3$ such that the
$D_r$ state is stably equivalent~\cite{cano2013bulk} to the $D_{r+8}$ theory up
to an additional $E_8$ state~\cite{KhanTeoHughesoappearsoon}, and both theories
have identical anyon fusion and braiding content. Taken together with the
non-simply laced $B_r$ series (which are non-Abelian at level-1 and will be
discussed elsewhere~\cite{KhanTeoHughesoappearsoon}), they form a class of
topological states with sixteenfold periodicity which matches the structure
found in Refs. \onlinecite{Kitaev06}, \onlinecite{LevinGu13}, and \onlinecite{fidkowski13a}.
Also, from the braiding phase and
 spin of the quasiparticles we expect
that  $K_{E_7} \oplus \sigma_x$ is stably equivalent to
$\left(-K_{su(2)}\right)\oplus E_8,$
$K_{E_6}\oplus\sigma_x\oplus\sigma_x$ is stably equivalent to
$\left(-K_{su(3)}\right)\oplus E_8,$ and $-K_{su(4)}\oplus E_8$ is stably equivalent to 
$ K_{so(10)}\oplus\sigma_x\oplus\sigma_x\oplus\sigma_x$.
%  where the over-bar refers to the anti-chiral time
% reversed partner.

% \begin{figure}[htbp]
% \centering
% \subfloat[Subfigure 1 list of figures text][Inner Automorphism where the lattice is rotated through $120^\circ$]{
% \includegraphics[width=0.17\textwidth]{su3latticsA}
% \label{fig:subfig1}}\quad
% \subfloat[Subfigure 2 list of figures text][Inner automorphism corresponding to reflection
% about the horizontal dotted grey line]{
% \includegraphics[width=0.17\textwidth]{su3latticsC}
% \label{fig:subfig2}}
% 
% \subfloat[Subfigure 3 list of figures text][Outer Automorphism where the lattice is rotated through $60^\circ$]{
% \includegraphics[width=0.17\textwidth]{su3latticsB}
% \label{fig:subfig3}}\quad
% \subfloat[Subfigure 4 list of figures text][Outer automorphism corresponding to reflection
% about the vertical dotted grey line]{
% \includegraphics[width=0.17\textwidth]{su3latticsD}
% \label{fig:subfig4}}
% \caption{Anyon lattice of $su(3)$ with inner and outer automorphisms. The white, red and 
% blue circles refer to the distinct quasiparticles $1$, $e$ and $e^2$ respectively 
% in the anyon quotient lattice 
% $\frac{\mathbb{Z}^2}{K\mathbb{Z}^2}$. 
% \subref{fig:subfig1} and \subref{fig:subfig2} are examples of inner automorphisms where anyon 
% labels (colors of the circles) are preserved whereas \subref{fig:subfig3} and \subref{fig:subfig4} 
% the outer automorphisms exchange $e$ and $e^2$ (blue and red circles).}
% \label{fig:su3lattice}
% \end{figure}
\begin{table}[htbp]
\centering
\begin{tabular}{lll}
&Anyon fusion lattice &\multirow{2}{*}{Anyon labels}\\&$\mathcal{A}=\mathbb{Z}^r/K\mathbb{Z}^r$&\\\hline
$A_r$&$\mathbb{Z}_{r+1}$&$1=e^{r+1},e,\ldots,e^r$\\
$D_{2n}$&$\mathbb{Z}_2\times\mathbb{Z}_2$&$1,e,m,\psi=e\times m$\\
$D_{2n+1}$&$\mathbb{Z}_4$&$1=e^4,e,e^2,e^3$\\
$E_6$&$\mathbb{Z}_3$&$1=e^3,e,e^2$\\
$E_7$&$\mathbb{Z}_2$&$1=e^2,e$\\
\end{tabular}
\caption{Quasiparticle labels of the $A_r=su(r+1)$, $D_r=so(2r)$ and $E_{6,7}$ Abelian topological states at level 1.}\label{tab:quasiparticles}
\end{table}

%%%%%%%%%%%%%%%% Anyonic symmetries %%%%%%%%%%%%%%%%

We will now construct the anyonic symmetry groups for the $ADE$ states and discuss applications of this result in the context of symmetry enhanced topological phases and semi-classical twist defects. One requirement of an AS is  that its operation commutes with the modular $S$ and $T$ transformations of topologically ordered (TO) states on a torus. For Abelian theories in the $K$-matrix formalism, a unitary anyon relabeling symmetry can be represented by a unimodular (integral entries, unit determinant) matrix $M$ that leaves the $K$-matrix invariant under $MKM^{T}.$ This forms a group of automorphisms \begin{align}\mbox{Aut}(K)=\left\{M\in GL(r;\mathbb{Z}):MKM^T=K\right\}\label{Autom}.\end{align} Since exchange and braiding are completely determined by the $K$-matrix,
the modular transformations are unchanged under the anyon relabeling,
\begin{align}S_{M{\bf a}M{\bf b}}=S_{{\bf a}{\bf b}},\quad T_{M{\bf a}M{\bf
b}}=T_{{\bf a}{\bf b}}\end{align} and the fusion rules remain unaltered as a
direct consequence of the linearity of $M$ or the Verlinde
formula~\cite{Verlinde88} in general.

Out of the full group of automorphisms there are certain trivial symmetry operations $M_0$ that only rearrange \emph{local} particles without changing the QP types. That is, these operations do not change the anyon equivalence classes  $[{\bf a}]={\bf a}+K\mathbb{Z}^r\in\mathcal{A}$ since they rotate the anyon lattice vector up to a local particle in $\Gamma=K\mathbb{Z}^r$. $M_0$ forms a normal subgroup of $\mbox{Aut}(K)$ called the {\em inner automorphisms} \begin{align}\mbox{Inner}(K)=\left\{M_0\in\mbox{Aut(K)}:[M_0{\bf a}]=[{\bf a}]\right\}.\end{align} To construct the relevant AS group we must remove this redundancy of trivial symmetry operations  by  quotienting, to generate the group known as the {\em outer automorphisms} \begin{align}\mbox{Outer}(K)=\frac{\mbox{Aut}(K)}{\mbox{Inner}(K)}.\label{outerdef}\end{align} Thus, $\mbox{Outer}(K)$ is the AS group of
the Abelian topological phase characterized by $K$. If the topological state is strictly charge conserving, a charge compatible AS element must keep the charge vector $M{\bf t}={\bf t}$ fixed which will ensure the charge of a QP is unchanged, $q_{M{\bf a}}=q_{\bf a}$. For a $U(1)$-breaking bosonic state at zero temperature, the charge compatibility condition can be relaxed modulo the image of $K$ so that the fractional charge is preserved by the symmetry only up to units of $e^\ast$ through the addition of local particles. Imposing these charge conservation conditions will further restrict the group of automorphisms in $\mbox{Aut}(K).$

\begin{figure}[t]
	\begin{center}
		\includegraphics[width=0.4\textwidth]{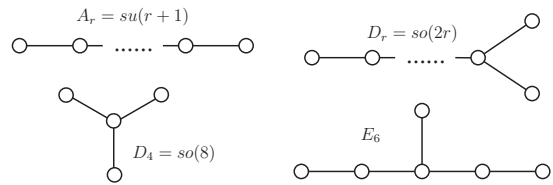}
	\end{center}
	\caption{Mirror symmetry of Dynkin Diagrams of $A_r,D_r,E_6$ and $S_3=Dih_3$ symmetry of $D_4$.}
	\label{fig:Dynkin}
\end{figure}

Remarkably, in an Abelian state in the $ADE$ classification, $\mbox{Outer}(K)$ is exactly 
the group of outer automorphisms of the
simply-laced Lie algebra, and coincides with the symmetry group of 
the Dynkin diagram~\cite{humphreys1972introduction,fuchs1995affine} (see Fig.~\ref{fig:Dynkin}). 
The explicit AS actions are listed in Table~\ref{tab:symmetries}. We present a brief discussion of the outer automorphisms of Lie algebras
 Appendix \ref{sec:Outer Automorphisms} where we motivate this connection in more mathematical detail. We omit $E_7$ from this list because
it has the anyon content \{$1, e$\}, $h_{1}=0;h_{e}=\frac{3}{4}$
(see Table \ref{tab:Tmatrixtable} ).
 Thus there are no anyonic symmetries for $E_7$.
For $A_2=su(3)$ the Dynkin diagram has a $\mathbb{Z}_2$ ``reflection" AS which is 
represented by the Pauli matrix $M_\sigma=\sigma_x$ that acts on the rank two anyon lattice vectors 
and simply interchanges the QPs $e$ and $e^2$ while leaving the vacuum fixed. 
Examples of explicit inner and outer automorphisms as represented on the anyon lattice are shown in Fig. \ref{fig:su3lattice}.

The only $ADE$ state with more than just a $\mathbb{Z}_2$ AS group is the $D_4=so(8)$ state which has a triality symmetry. The AS group is $S_3=Dih_3$ which is the permutation group of three elements generated by ``reflection" and threefold ``rotation" in its Dynkin diagram. This group is non-Abelian, contains a total of six elements, and the generators
are represented by \begin{align}M_{\sigma_\psi}=\left(\begin{array}{*{20}c}1&0&0&0\\0&1&0&0\\0&0&0&1\\0&0&1&0\end{array}\right),\quad
M_{\rho}=\left(\begin{array}{*{20}c}1&0&0&0\\0&0&0&1\\0&1&0&0\\0&0&1&0\end{array}\right)\label{eq:so8generators}\end{align} which act on four dimensional anyon lattice vectors. $M_{\sigma_\psi}$ interchanges $e\leftrightarrow m$ but fixes $\psi$ up to local boson. $M_\rho$ rotates $e\to m\to\psi\to e$ which is an example of a threefold symmetry operation. All $ADE$ symmetry operations can be chosen to strictly preserve $U(1)$ symmetry and leave the charge vector unchanged so we have no further restrictions. In fact, to our knowledge this is the only known example of a non-Abelian anyonic symmetry group.

\begin{table}[t]
\centering
\begin{tabular}{lll}
&$\mbox{Outer}(K)$&anyonic symmetry action\\\hline
$A_r,D_{2n+1},E_6$&$\mathbb{Z}_2$&$\sigma:e\leftrightarrow e^{-1}$\\
$D_{2n}$, $n\neq2$&$\mathbb{Z}_2$&$\sigma:e\leftrightarrow m$\\
\multirow{2}{*}{$D_4$}&\multirow{2}{*}{$S_3=Dih_3$}&$\sigma_\psi:e\leftrightarrow m$\\&&$\rho:e\to m\to\psi\to e$
\end{tabular}
\caption{Symmetry action $\mbox{Outer}(K)$ on anyon labels.}\label{tab:symmetries}
\end{table}

Now that we have introduced the basic idea behind AS we will move on to discuss the consequences. The remainder of the article is organized as follows: (i) we first show/review that the generic consequence of a non-trivial AS group is the existence of  topologically distinct gapped
phases along quasi-one dimensional interfaces/edges with identical, 
but oppositely propagating, edge modes which chiral central charge $c_{-}=0$ (see
Fig.~\ref{fig:edgecouplings})\cite{BarkeshliJianQi13long}; (ii) since counter-propagating modes at an interface can be mapped to a single edge of a fractional quantum spin Hall phase we can apply our results to determine the stability of the edges of ADE FQSH states; (iii) finally we will use the AS group to determine the distinct set of semiclassical twist defects in ADE FQH states and some of their properties.

%.............put to introduction..............
%Recently there has been a lot of excitement about anyonic symmetries in an Abelian system and extrinsic twist defects hosting nonAbelian quasiparticles. Anyon symmetries act by

%relabelling a set of anyons such that their spin and braiding is left invariant. Examples include the exchange of $e$ and $m$ in toric code\cite{KitaevKong12,Bombin} and interchanging
%layer labels in the bilayer quantum hall state \cite{BarkeshliQi,BarkeshliJianQi,teo2013braiding}. Other examples include quantum rotor models \cite{YouWen} and $S_3$ symmetry enhanced
%models \cite{TeoRoyXiao13long}.

%toric code, bilayer quantum Hall states, color code
%time reversal symmetric topologically ordered surface state

%put detail computation of Aut(K) and Inner(K) to supplementary

%%%%%%%%%%%%% Weakly coupled interfaces %%%%%%%%%%%%%%%%%%%%%%%%%%

\section{Anyonic symmetry and distinct Gapped Interface Phases }

\begin{figure}[htbp]
\centering
\includegraphics[width=0.4\textwidth]{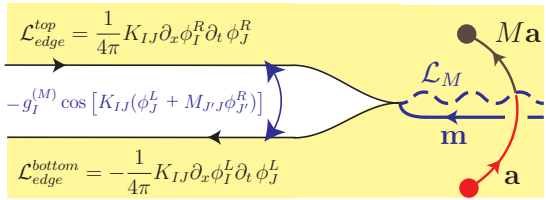}
\caption{Local boson tunneling in \eqref{edgecoupling} gives a gapped interface phase $\mathcal{L}_M$ represented by a branch cut (dashed
wavy blue line) and a parallel quasiparticle string (solid blue line). Passing anyon changes type ${\bf a}\to M{\bf a}$ and accumulate a crossing phase $\mathcal{D}S_{{\bf a}{\bf m}}$ in
\eqref{condensate}.}\label{fig:edgecouplings}
%\includegraphics[width=0.25\textwidth]{defect.eps}
%\caption{Representation of twist defect (red cross) dressed with branch cuts (dashed wavy lines) and quasiparticle strings (solid lines).}\label{fig:defect}
\end{figure}

We begin our study with a quasi-one dimensional interface with identical but oppositely propagating edge modes.
The gapless edge modes on an interface have the $(1+1)$-d bosonic Lagrangian density
$\mathcal{L}_{edge}=\frac{1}{4\pi}K_{IJ}\partial_x\phi_I\partial_t\phi_J$, where
the $K$-matrix is identical to that in the bulk, and QPs are expressed as vertex
operators $\psi^{\bf a}=e^{i{\bf a}\cdot\boldsymbol{\phi}}$. 
In limit of weak coupling between opposite sides of the interface, the chiral gapless edge modes along the opposite sides of the interface in question are
described by the boson Lagrangian density
\begin{align}\mathcal{L}^{top}_{edge}+\mathcal{L}^{bottom}_{edge}=\frac{1}{4\pi}
K^{\sigma\sigma'}_{IJ}\partial_x\phi_I^\sigma\partial_t\phi_J^{\sigma'}{
+\frac{1}{2\pi}t^{\sigma}_{I}\epsilon^{\mu\nu}\partial_{\mu}\phi^{\
\sigma}_IA_{\nu}}  \label{interfaceK}\end{align} where $\sigma=0,1=R,L$
labels right and left moving modes, $\phi^R_I$ ($\phi^L_I$) are the boson fields living along
the top (bottom) edge, and
$K^{\sigma\sigma'}_{IJ}=(-1)^\sigma\delta^{\sigma\sigma'}K_{IJ}$ (see Fig. \ref{fig:edgecouplings}).
We assume that
$\mathbf{t}^\sigma=\mathbf{t}^L=\mathbf{t}^R=\mathbf{t}$. Corresponding to each
quasiparticle vector $\mathbf{a}$ in the bulk, the vertex operator on the
right(left) edge is
$\psi_R^{\mathbf{a}}=e^{i\mathbf{a}\cdot\boldsymbol{\phi}^{R}}
(\psi_L^{\mathbf{a}}=e^{-i\mathbf{a}\cdot \boldsymbol{\phi}  ^ { L } } )$.
This convention ensures that the
charge of both
$\psi_R^{\mathbf{a}}$
and $\psi_L^{\mathbf{a}}$ is $e^{\ast}\mathbf{a}^T K^{-1}\mathbf{t}$ .

Remarkably, given any element $M$ of the AS group, the interface can be
gapped by a corresponding set of backscattering terms
\begin{align}\delta\mathcal{L}_M=-\sum_I g^{(M)}_I\cos\left[K_{IJ}
\left(\phi^L_J+M_ {
J'J }\phi^R_{J'}\right)\right]\label{edgecoupling}\end{align} where repeated
indices $J$ and $J'$ are summed over. This describes tunneling between the local
boson
$e^{iK_{IJ}M_{J'J}\phi^R_{J'}}$ on the top edge and
$e^{-iK_{IJ}\phi_J^L}$ on the bottom. We assume that the representative matrix $M$ of the AS group element is charge
conserving so that the tunneling term preserves boson number, however, this condition can
be relaxed if the global $U(1)$ charge symmetry is broken as discussed earlier.

In the strong-coupling limit the terms in Eq.~\eqref{edgecoupling} lead to a
gapped interface phase associated with the symmetry $M$ which we denote by the symbol $\mathcal{L}_M$ for convenience. The backscattering terms simultaneously pin the
boson vacuum expectation values
\begin{align}\langle\phi^L_I+M_{JI}\phi^R_J\rangle=2\pi(K^{-1})_{
IJ } m_J , \quad\mbox{for ${\bf
m}\in\mathbb{Z}^r$}\label{eq:bosonpinning}\end{align} as the operators for each
value of $I$ mutually commute. This pinning effectively
condenses the local bosonic QP pairs 
\begin{align}
\left\langle\left(\psi^{M\mathbf{a}}_R\right)^\dagger\psi^{\bf a}_L\right\rangle=\langle e^{-i({\bf
a}\cdot\boldsymbol{\phi}^L+(M{\bf a})\cdot\boldsymbol{\phi}^R)}\rangle=e^{-2\pi i{\bf
a}^TK^{-1}{\bf m}}\label{condensate}
\end{align}\noindent along the interface. The gapped interface state is then characterized by this
QP pair condensation, and can be diagrammatically represented by a branch cut associated with 
$M$ that is decorated with a parallel
QP string ${\bf m}$ localized near the cut (see Fig.~\ref{fig:edgecouplings}). The branch cut itself will change the anyon type of a passing QP from ${\bf a}\to M{\bf a},$ while the attached ${\bf m}$ string contributes the additional $U(1)$ crossing phase required from  Eq. \eqref{condensate}. Such QPs localized at defects have also been studied by in Refs. \onlinecite{BarkeshliJianQi13long, LindnerBergRefaelStern, ClarkeAliceaKirill, MChen, Vaezi}. %However the situation here is somewhat different because we do not consider superconductivity. 
Two anyonic symmetry matrices $M$ and $M'$ correspond the same gapped interface phase if the backscattering terms pin and condense
the same set of bosonic QP pairs, i.e., the gapped edge phases are identical if $M{\bf a}=M'{\bf a}$ modulo $K\mathbb{Z}^r$. Gapped interface phases are therefore in one-to-one correspondence to the group $\mbox{Outer}(K)$ of anyonic relabeling symmetries defined in \eqref{outerdef}.  We further note that, although we focus on charge conserving
edge tunneling terms in this work, our formalism also applies to gapping terms which
describe superconducting pairing. This includes terms like
\begin{align*}
 \delta\mathcal{L}_M&=-\sum_I g^{(M)}_I
e^{iK_{IJ}M_{J'J'}\phi^R_J}e^{-iK_{IJ}\phi^{L}_J}+h.c.\\
&=-\sum_I g^{(M)}_I\cos\left[K_{IJ}
\left(\phi^L_J-M_ {
J'J }\phi^R_{J'}\right)\right]
\end{align*} 
which condense local bosonic pairs
$\psi^{M\boldsymbol{a}}_R\psi^{\boldsymbol{a}}_L$ on the edge.

\section{Application I: Stability of ADE bosonic fractional quantum spin Hall states}
%%%%%%%%%%%%%%%%%%%%%%%%%%%%
%Now, with this setup we are equipped to consider two interesting problems: the gapped edge phases of
%$ADE$ fractional quantum spin Hall phases and $ADE$ twist defects.

In this section we argue that the Lagrangian for the interface is identical to that of the edge 
of a bosonic quantum spin hall state.
This enables us to classify gapped edges of bosonic FQSH using anyonic symmetries. We will outline the action of time reversal operators on the edge degrees of the 
quantum spin hall state (with more details in the appendix). To understand the stability of the FQSH edge states we  must carefully analyze whether the gapping terms we add to destabilize the gapless degrees of freedom
on the edge explicitly or spontaneously break TR symmetry. Such questions have been the focus of recent works like \cite{LevinStern12,LevinWang}. 
Once we understand the symmetry properties of the possible gapped edge phases we can determine the conditions under which the edge can be destabilized without breaking any protective symmetries.

To approach the analysis, we first note that the two edge theories coupled at an interface in the previous section (c.f. Eq. \eqref{interfaceK}) can be regarded as a single-edge of a doubled system with 
$K$-matrix $K^{\sigma\sigma'}=K\oplus(-K)$ if the topological states on the two sides of the interface are folded on top of each 
other~\cite{BarkeshliJianQi13, BarkeshliJianQi13long}. The $K$ matrix of a such a theory looks like
$
\begin{pmatrix}
 K&0\\
 0&-K
\end{pmatrix} 
$ with bosonic degrees of freedom on the edge represented by 
$\begin{pmatrix}
  \phi_R\\
  \phi_L
 \end{pmatrix}
$, where $\phi_R$ and $\phi_L$ are counterpropagating edge modes on the same interface. We see that this Lagrangian is the same as \eqref{interfaceK}.
This represents a fractional quantum spin Hall state of 
bosons~\cite{LevinStern09, LevinStern12}. This identification enables us to characterize the gapped edges of these states using anyonic symmetry. The time reversal (TR) matrix is $T^{\sigma\sigma'}_{IJ}=(\sigma_x)^{\sigma\sigma'}\delta_{IJ}$ 
acting on the spin-momentum locked $\sigma=\uparrow,\downarrow=R,L$ degree of freedom. The
TR operator $\mathcal{T}$ is anti-unitary and acts according to 
\begin{align}
\mathcal{T}^{-1}\phi^\sigma_I\mathcal{T}=T^{\sigma\sigma'}_{IJ}\phi^{\sigma'}_J+\pi(K^{-1})^{\sigma\sigma'}_{IJ}\chi^{\sigma'}_J\label{TRdef}
\end{align}
for some TR vector $\boldsymbol{\chi}=(\boldsymbol{\chi}_\uparrow,\boldsymbol{\chi}_\downarrow)\in\mathbb{R}^{2r}.$\cite{LevinStern12}  For 
our classes of  ADE FQSH systems (and many others) we show in Appendix {\ref{subsec:chizero}}   that 
$\boldsymbol{\chi}$ can be set to $0$ and hence, $\mathcal{T}^{-1}\phi^{L/R}_{I}\mathcal{T}=\phi^{R/L}_{I}$. Thus,  $\,\,\mathcal{T}^{-1}\psi_R^{\mathbf{a}}\mathcal{T}=\psi_L^{\mathbf{a}}$. Indeed since $\psi^{\mathbf{a}}_R$ and 
$\psi_L^{\mathbf{a}}$ are time-reversed partners, their T and S
matrices obey
$\theta_R^{\mathbf{a}}=\left(\theta_L^{\mathbf{a}}\right)^{\ast}=e^{i\pi\mathbf{a}^TK^{-1}\mathbf{b}}$ and
$S_R^{\mathbf{ab}}=\left(S_L^{\mathbf{ab}}\right)^{\ast}=e^{i\pi\mathbf{a}^TK^{-1}\mathbf{b}}.$
Also, under time reversal the gapping term transforms as
\begin{eqnarray*}
\delta\mathcal{L}_{TM}&=&\mathcal{T}^{-1}\delta\mathcal{L}_M\mathcal{T}\nonumber\\
&=&-\sum_I g^{(M)}_I\cos\left[K_{IJ}
\left(\phi^R_J+M_ {
J'J }\phi^L_{J'}\right)\right].
\end{eqnarray*}

Generically, since the edge is non-chiral, one can destabilize the edge and open a gap via, for example, condensing bosons on the edge. However, we are not only interested if a gap can form, but what symmetries the resulting gapped state breaks or preserves, e.g., some gapped phases may break time-reversal and some may preserve time-reversal.  For our problem, the edge condensate is formed from 
the QP pairs $\left(\psi^{M\bf a}_{\uparrow}\right)^{\dagger}\psi^{{\bf a}}_{\downarrow}$, and is a maximal collection of mutually local bosons, 
known as a {\em Lagrangian subgroup}~\cite{Levin13} in the TR symmetric doubled anyon lattice system
$\mathcal{A}_{\uparrow}\otimes\mathcal{A}_{\downarrow}$. 

\subsection{Explicit T-Breaking on the edge of  bosonic FQSH systems}
Let us look at a few examples before making a general statement. To simplify notation let $x_{R/L}$ stand for $\psi^{\mathbf{x}}_{R/L}$ and 
$\bar{x}_{R/L}\equiv\left(\psi^{\mathbf{x}}_{R/L}\right)^{\dagger}$. The $A_2=su(3)$ state has AS $\mbox{Outer}(K_{su(3)})=\mathbb{Z}_2$ generated by $1$ and $\sigma$.
These generators correspond to the gapped interface/edge phases $\mathcal{L}_1$ and $\mathcal{L}_\sigma$ with QP pair 
condensates $\{1_L\overline{1}_R,e_L\overline{e}_R,e^2_L\overline{e^2}_R\}$ and $\{1_L\overline{1}_R,e_L\overline{e^2}_R,e^2_L\overline{e}_R\}$ respectively. To test if these phases break TR explicitly we calculate
$\mathcal{T}^{-1}\mathcal{L}_1\mathcal{T}=\{1_R\overline{1}_L,e_R\overline{e}_L,e^2_R\overline{e^2}_L\}$ and 
$\mathcal{T}^{-1}\mathcal{L}_\sigma\mathcal{T}=\{1_R\overline{1}_L,e_R\overline{e^2}_L,e^2_R\overline{e}_L\}$.
To compare with $\mathcal{L}_1$ and $\mathcal{L}_\sigma$ we take the Hermitian conjugate, and we see that
$\mathcal{T}^{-1}\mathcal{L}_{1}\mathcal{T}=\left(\mathcal{L}_{1}\right)^{
\dagger}$ and
$\mathcal{T}^{-1}\mathcal{L}_{\sigma}\mathcal{T}=\left(\mathcal{L}_{\sigma}
\right)^{\dagger}$. While this might initially seem problematic for TR preservation, we note that $\mathcal{L}$ and
$\mathcal{L^{\dagger}}$ actually represent the same set of condensed bosons.  We see
by
using equations \eqref{eq:bosonpinning} and \eqref{condensate} that
\begin{align*}
 \left\langle\left(\psi^{M\mathbf{a}}_R\right)^\dagger\psi^{\bf
a}_L\right\rangle\neq 0\implies 
\left\langle\psi^{M\mathbf{a}}
_R\left(\psi^{\bf
a}_L\right)^{\dagger}\right\rangle\neq 0.
\end{align*}
Explicitly,  we can see this by
taking the Hermitian conjugate of equation \eqref{condensate}
\begin{align*}
\left\langle\left[\left(\psi^{M\mathbf{a}}_R\right)^{\dagger}\psi^{\bf
a}_L\right]^{\dagger}\right\rangle&=\left\langle\psi^{M\mathbf{a}}
_R\left(\psi^{\bf
a}_L\right)^{\dagger}\right\rangle=e^{2\pi
i{\bf
a}^TK^{-1}{\bf m}}.
\end{align*}
Thus, we see that we can safely add both $\delta\mathcal{L}_M$ and $\delta\mathcal{L}_{TM}$ (i.e., the terms corresponding to the time-reversed partner of the $\delta\mathcal{L}_m$)
to the edge Lagrangian  $\mathcal{L}^{top}_{edge}+\mathcal{L}^{bottom}_{edge}$ in this case{(and in all cases with $\mathbb{Z}_2$ anyonic symmetry as we show in the next subsection)}
{\emph{while staying in the same phase}} (that is condensing the same set of bosons) .
Hence, $\mathcal{L}^{top}_{edge}+\mathcal{L}^{bottom}_{edge}+\delta\mathcal{L}_M+\delta\mathcal{L}_{TM}$ is TRI. 
Thus, we see that when the time reversed (i.e., Hermitian conjugate)
phase condenses the same bosons  as the original one, i.e.  it does not break time reversal
{\emph{explicitly}}.
These statements pertain to the explicit breaking of time-reversal. We
will deal with spontaneous symmetry breaking below where we check  that the expectation value of the condensate and its TR partner are the same. As we will show the condition for the absence of spontaneous time-reversal breaking will be
Eq. \eqref{eq:TRSYMMETRYCONDITION}. Before we discuss this, let us explicitly consider the phases of the $so(8)$ state.

The $D_4=so(8)$ state has six symmetry operators in $\mbox{Outer}(K_{so(8)})=S_3$,
and these correspond to six gapped phases. For example the trivial one $\mathcal{L}_1$ condenses 
$\{1_L\overline{1}_R,e_L\overline{e}_R,m_L\overline{m}_R,\psi_L\overline{\psi}_R\}$, the twofold one 
$\mathcal{L}_{\sigma_\psi}$ condenses $\{1_L\overline{1}_R,e_L\overline{m}_R,m_L\overline{e}_R,\psi_L\overline{\psi}_R\}$, 
and the threefold one $\mathcal{L}_{\rho}$ condenses $\{1_L\overline{1}_R,e_L\overline{m}_R,m_L\overline{\psi}_R,\psi_L\overline{e}_R\}$. 
The Lagrangian subgroup for $\mathcal{L}_{\sigma_\psi}$ preserves TR (analogous to $\mathcal{L}_\sigma$ for $su(3)$) while that for $\mathcal{L}_\rho$ breaks it
upon adding the gapping terms. We consider the example for $\mathcal{L}_{\rho}$ explicitly,
$\mathcal{T}^{-1}\mathcal{L}_{\rho}\mathcal{T}=\{1_R\overline{1}_L,e_R\overline{
m}_L,m_R\overline{\psi}_L,\psi_R\overline{e}_L\}$. Taking hermitian conjugate we
see that
$\{1_L\overline{1}_R,m_L\overline{e}_R,\psi_L\overline{m}_R,e_L\overline{\psi}
_R\}=\mathcal{L}_{\rho^{-1}}\neq \mathcal{L}_{\rho}$. Hence the edge phase determined by the threefold symmetry \emph{explicitly} breaks time
reversal symmetry, because the time-reversed phase condenses a different set of bosons.

After the intuition gained from examining these cases we will now prove that the gapped edge phase $\mathcal{L}_M$
 does not break TR explicitly
if the Lagrangian subgroup is TR invariant, i.e., the TR of the gapped
edge describes the same phase, which is the case if and only if $M^2=1$ (up to
inner automorphisms). Microscopically this arises from the fact that if $M^2=1$ then the sine-Gordon gapping terms in \eqref{edgecoupling} can be made TR symmetric by adding time-reversed counterparts 
\emph{that pin the same set of QP pairs}. This includes all symmetries in the $ADE$ states except the threefold symmetry $\rho$ of $so(8)$. One cannot write down a time-reversal invariant Lagrangian that is gapped by the three-fold symmetry group element.

% However, \cite{LevinStern09,LevinStern12} have a very simple criterion which allows 
% us to know the answer immediately.\\
% Let $q_{\text{min}}$ be the minimum charge among all excitations of the anyon 
% lattice of $K^{\sigma\sigma'}$(ignoring the vacuum sector). If 
% \begin{equation*}
%   \frac{\left( \boldsymbol{\chi^\sigma} \right)^T{\left( K^{\sigma\sigma'} \right)}^{-1}\left( \mathbf{t}^\sigma \right)}{q_{\text{min}}}=1\quad\text{mod }2
% \end{equation*}
% only then do we have a protected state. However, as showed in the previous section for us $\boldsymbol{\chi}^\sigma=\mathbf{0}$.
% Thus, \emph{for any doubled bosonic system with oppositely propagating edge modes we can always gap out the edge 
% without breaking either time reversal or charge conservation.} Hence, in all the $ADE$ systems studied in the main text
% we should have no protected ege modes.
% 
% Now that we have stated the result, let us explicitly write down the gapping terms and show that
% they indeed do not break TRI.\\
We can continue our microscopic treatment by expanding the edge Lagrangian density in \eqref{interfaceK},
\begin{align*}
  \mathcal{L}_{\text{edge,bare}}&=\frac{1}{4\pi}K_{IJ}\partial_x\phi_I^R\partial_t\phi_J^R-\frac{1}{4\pi}K_{IJ}\partial_x\phi_I^L\partial_t\phi_J^L\\&+
  \frac{e^\ast}{2\pi}\epsilon  ^{\mu\nu}t_I\partial_{\mu}\phi_{I}^RA_{\nu}
  +\frac{e^\ast}{2\pi}\epsilon  ^{\mu\nu}t_I\partial_{\mu}\phi_{I}^LA_{\nu}.
\end{align*}
$\mathcal{L}_{\text{edge}}$ is obviously TRI. But, now we need to understand what happens when we add the gapping term $\delta\mathcal{L}_M$ in \eqref{edgecoupling}  ; we reproduce it here for convenience
\begin{align*}
 \delta\mathcal{L}_M=-\sum_Ig^{(M)}_I\cos\left[K_{IJ}\left(\phi^L_J+M_ {
J'J }\phi^R_{J'}\right)\right].
\end{align*}
$\delta\mathcal{L}_M$  breaks TRI, to restore it we add its time reversed partner
\begin{align*}
   \delta\mathcal{L}_{TM}&=\mathcal{T}^{-1}\left(\delta\mathcal{L}_M\right)\mathcal{T}\nonumber\\&=-\sum_Ig^{(M)}_I\cos\left[K_{IJ}\left(\phi^R_J+M_ {
J'J }\phi^L_{J'}\right)\right].
\end{align*}
The full Lagrangian $ \mathcal{L}_{\text{edge,bare}}+ \delta\mathcal{L}_M+ \delta\mathcal{L}_{TM}$ is time reversal invariant, but we also need to make sure that \emph{$\delta\mathcal{L}_M$ and $\delta\mathcal{L}_{TM}$ describe the same gapped phase, i.e. both the terms condense the same set of bosonic quasiparticles.}
This is equivalent to the statement that
\begin{align*}
K_{IJ}\left(\phi^R_J+M_{J'J}\phi^{L}_{J'}\right)&=2\pi\left(p_1\right)_I;\quad \left(p_1\right)_I\in \mathbb{Z}\\
\implies K_{IJ}\left(\phi^L_J+M_{J'J}\phi^{R}_{J'}\right)&=2\pi\left(p_2\right)_I;\quad \left(p_2\right)_I\in \mathbb{Z}.
\end{align*}
Next we find the conditions when the above is true (we use vector notation from here on and $K$ and $M$ are matrices)
\begin{align}
K\left(\boldsymbol{\phi}^R+M^T\boldsymbol{\phi}^L\right)&=2\pi\mathbf{p_1}\end{align}
and using $MKM^T=K$ we see
\begin{align}
K\left(\boldsymbol{\phi}^L+\left(M^T\right)^{-1}\boldsymbol{\phi}^R\right)&=2\pi M\mathbf{p_1}.
\label{eq:TRtrap}
\end{align}
Now, drawing upon the correspondence between gapped interface phases and AS outlined before, this implies that if $\delta\mathcal{L}_M$ leads to the gapped interface which acts on the anyons labels $\mathbf{a}\rightarrow M\mathbf{a}$,  $\delta\mathcal{L}_{TM}$ leads to $\mathbf{a}\rightarrow M^{-1}\mathbf{a}$. Further, as remarked before, 
two anyon symmetries $M$ and $M'$ lead to the same gapped interface if $M\mathbf{a}=M'\mathbf{a}+K\mathbb{Z}^r\,\forall \mathbf{a}$, which is to say $M=M'$ up to inner automorphisms. In this case we require $M=M^{-1}$, thus $M^2=1$ up to inner automorphisms.

\emph{Thus if the gapped edge of the bosonic quantum spin hall effect is in the phase $\mathcal{L}_M$ and preserves TRI it must satisfy $M^2=1$}. We, have already seen the example of the three fold symmetry $\rho$ of $so(8)$, which acts on the anyon labels by sending
$(e,m,\psi)\rightarrow (m,\psi,e)$ . It satisfies $\rho^3=1$. $\delta\mathcal{L}_{T\rho}$ however represents
the inverse threefold symmetry $\rho^{-1}$, by equation \eqref{eq:TRtrap}. Since $\rho^{-1}\neq \rho$ this breaks TRI explicitly as mentioned above. 

\subsection{Spontaneous T-breaking on the edge of Bosonic FQSH systems}
This concludes our discussion of the explicit breakdown of TRI.
There is a further complication in that, even though the Lagrangian is TR invariant, the ground state condensate itself can break TR spontaneously. As such,  even when a gapped interface may pin the
same QPs as its TR partner, the expectation value of the condensate in the time reversed phase
may be different than that in the original phase. In order to prevent the spontaneous breaking of TR symmetry we must  constrain the condensate phase as follows. First, let us specialize to gapped edges where $M^2=1$ so that TR is not broken explicitly.
Time reversal \eqref{TRdef} operates on the QP condensate $\left(\psi^{M\bf
a}_{R}\right)^\dagger\psi^{\bf a}_{L}$ 
along a fractional quantum spin Hall edge by 
\begin{align}
\mathcal{T}^{-1}\left[\left(\psi^{M\bf a}_{R}\right)^\dagger\psi^{\bf
a}_{L}\right]\mathcal{T}&=
\mathcal{T}^{-1}\left[e^{-i((M{\bf a})\cdot\boldsymbol\phi^R+{\bf
a}\cdot\boldsymbol\phi^L)}\right]\mathcal{T}
\nonumber\\
=e^{i((M{\bf a})\cdot\boldsymbol\phi^L+{\bf
a}\cdot\boldsymbol\phi^R)}&=e^{i((MM{\bf a})\cdot\boldsymbol\phi^R+{M\bf
a}\cdot\boldsymbol\phi^L)}.\nonumber
\end{align} 
Using Eq. \eqref{eq:bosonpinning} we find
\begin{align}
\left\langle\mathcal{T}^{-1}[\left(\psi^{M\bf a}_{R}\right)^\dagger\psi^{\bf
a}_{L}]\mathcal{T}\right\rangle&=e^{
2\pi i{\bf a}^TK^{-1}(M{\bf m})},\nonumber 
\end{align} which should be compared to $\left\langle\left(\psi^{M\bf a}_{R}\right)^\dagger\psi^{\bf
a}_{L}\right\rangle=e^{-2\pi i{\bf a}^TK^{-1}{\bf
m}}$ from Eq. \eqref{condensate}.
Thus, to preserve time-reversal we must have ${\bf m}$ 
satisfy 
\begin{align}{\bf m}\equiv -M{\bf m}\quad\mbox{mod $K {\bf m}$}
\label{eq:TRSYMMETRYCONDITION}
\end{align} 
otherwise the ground state will break TR \emph{spontaneously}. For all of the $ADE$
cases, we have examples of $M$ which obey 
$M^2=1,$  and conserve charge $M\mathbf{t}=\mathbf{t}$
(Eqs.~\eqref{eq:so8generators} and \eqref{sigmaADE}). 
We can also satisfy Eq. \eqref{eq:TRSYMMETRYCONDITION} with the case
when $m=0$.  \emph{Thus in all the $ADE$ FQSH cases, we can gap out the edge even if we
demand TRI and charge conservation}. 

In fact, using the criterion in Ref. 
\onlinecite{LevinStern12}, one can show in general that whenever $\boldsymbol{\chi}=0$, the edge
can be gapped without breaking TR or charge conservation.
In particular these gapped edges represent twofold defects $M^2=1$. However,
there exist gapped edges which do break time reversal symmetry explicitly while
conserving charge as well, e.g.,  the threefold defect $\rho$ for $so(8)$.

%\eqref{eq:TRSYMMETRYCONDITION} in the appendix.

%%%%%%%%%%%%%%%%%%% Twist Defect %%%%%%%%%%%%%%%%%%%%%%%%%%
\section{Application II: $ADE$ Twist Defects from Anyonic Symmetry}
\begin{figure}[t]
\centering
\includegraphics[width=0.35\textwidth]{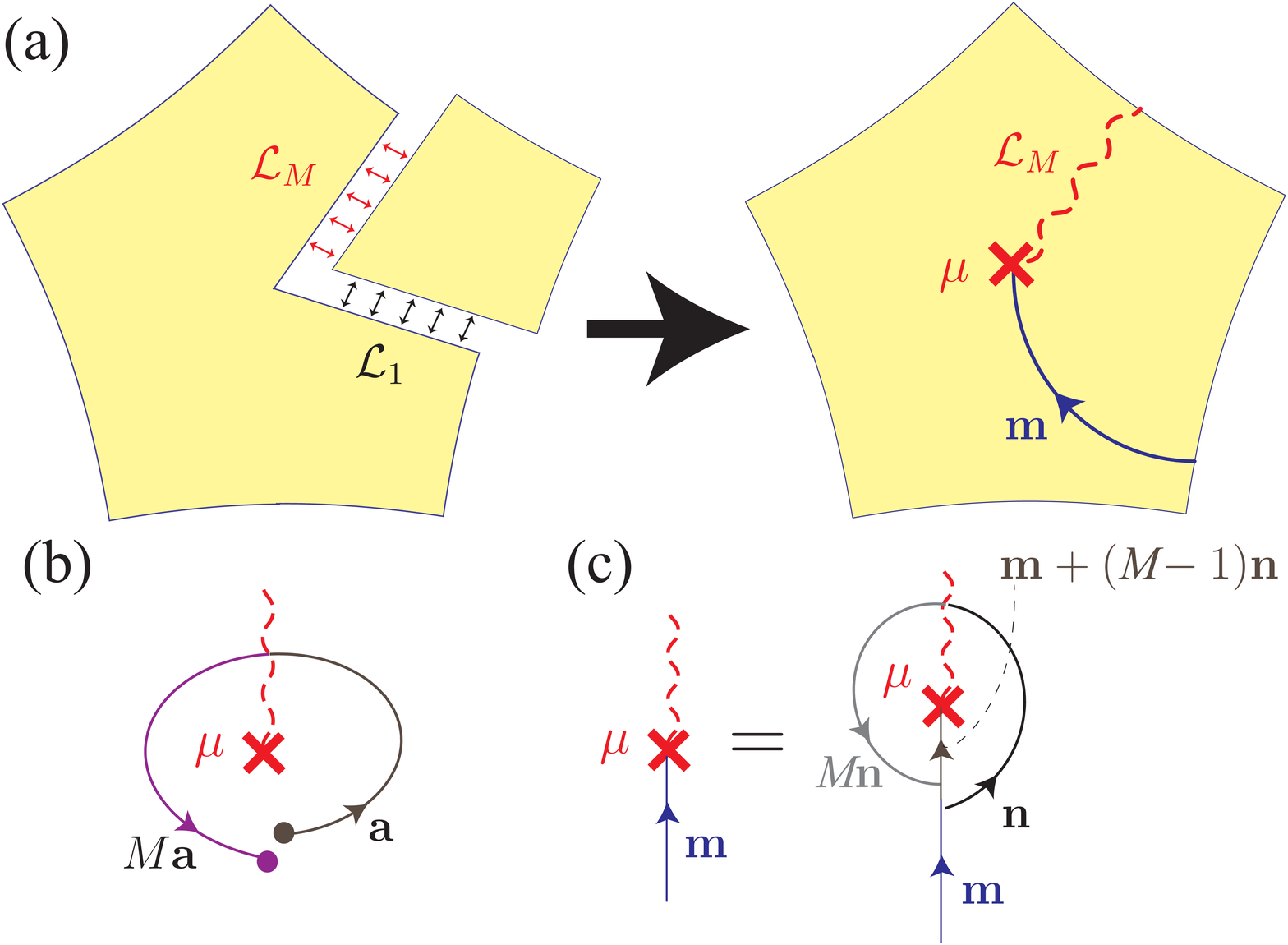}
\caption{(a) Twist defect $\mu$ (red cross) as a disclination at the end of a gapped interface $\mathcal{L}_M$ (dashed curvy line) and dressed with a quasiparticle string ${\bf m}$ (solid line). (b) Defect action on a circling quasiparticle ${\bf a}\to M{\bf a}$. (c) Redundancy in defect-quasiparticle fusion.}\label{fig:volterra}
\end{figure}

In addition to gapping symmetry protected edge states, we can apply our results to classify twist defects associated with an AS element $M$, i.e., a topological point defect that changes the anyon type of QPs that travel around it according to  ${\bf a}\to M{\bf a}$ (see Fig.~\ref{fig:volterra}b). As discussed above, twist defects can essentially be considered as a domain wall sandwiched between two distinct gapped interface phases~\cite{BarkeshliJianQi13long}. They can also appear attached to dislocations or disclinations (see Fig.~\ref{fig:volterra}a) where non-trivial boundary conditions (i.e. local boson tunneling) are applied at the extra inserted half-layer or wedge respectively. The classical defect
interpretations are particularly relevant when the TO intertwines with (liquid) crystalline order where a broken discrete spatial symmetry matches an AS~\cite{Kitaev97, Wenplaquettemodel, BombinColorCode, TeoRoyXiao13long}.

According to Eq.~\eqref{condensate} a twist defect $\mu$ can bind a QP ${\bf m}$ (see Fig.~\ref{fig:volterra}a,c). The defect-QP composite is summarized by a defect {\em species} label $\lambda$ where $\mu_\lambda\equiv\mu_0\times{\bf m}$, where $\mu_0$ is a {\em bare} defect with no attached QP, and $\times$ means fusion. Because of the intrinsic nature of the twist defect, there is an important consistency constraint that must be satisfied when determining the possible attached QP types, i.e., there are redundancies in defect-quasiparticle fusion. For instance, as shown in Fig.~\ref{fig:volterra}c, when the QP ${\bf m}$ is fusing with the defect $\mu$ it can emit a QP ${\bf n}$ that travels around the twist defect and is re-absorbed after (possibly) changing its anyon type. If the twist defect does not transform ${\bf n}$ then it is simply re-absorbed without issue, however if ${\bf n}$ is transformed then a physical consistency condition must be satisfied that leads to a redundancy in the possible defect 
species 
types.  The constraint that must be satisfied is \begin{align}\mu\times{\bf m}=\mu\times\left({\bf m}+(M-1){\bf n}\right)\end{align} and the defect species labels are thus classified by equivalence classes $\lambda=\llbracket{\bf m}\rrbracket\equiv[m]$ modulo $(M-1)\mathcal{A}$~\cite{teo2013braiding}. A heuristic way to understand this constraint is that if the emitted quasi-particle ${\bf n}$ is changed to $M{\bf n}$ by the twist defect then the defect itself must be able to absorb the difference between ${\bf n}$ and $M{\bf n}$ to give a physically consistent result. As we will see, this means that the defects themselves can be thought of heuristically to  ``contain" internal structure that increases their quantum dimension and allows them to compensate for this emission-reabsorption process.

Take the the $A_r=su(r+1)$ state for example. By solving the redundancy constraint  we can see that the non-redundant species label only represents the parity of an $e^p$ QP  bound to a twofold defect $\sigma.$ However, the  parity (i.e., even or oddness of $p$) is only well defined only when the particular $A_r$ has well defined even and odd QP sectors, i.e. when there are even number of QPs, i.e., when $r$ is odd. Thus $A_r$, when $r$ is even, only has one type of defect species (equivalent to the bare defect), and when $r$ is odd there are two types, an even and odd defect $\sigma_0$ and $\sigma_1.$ It is interesting to note that when $r$ is even the defect internally harbors all of the distinct QPs leading to a quantum dimension of $\sqrt{r+1}$ while for $r$ odd the burden is shared between the two different defect species which each carry $\sqrt{(r+1)/2}.$
%Take the the $A_r=su(r+1)$ state for example. The operation $1-\sigma$ sends the quasiparticle $e$ to $e^2$. This is invertible for even $r$ when the anyon lattice $\mathcal{A}_{su(r+1)}=\mathbb{Z}_{r+1}$ has an odd order. Defect species labels live in the quotient $\mathcal{A}_{su(r+1)}/(1-\sigma)\mathcal{A}_{su(r+1)}=\mathbb{Z}_{r+1}/2\mathbb{Z}_{r+1}$, which is trivial when $r$ is even or $\mathbb{Z}_2$ when $r$ is odd. 
For the $D_4=so(8)$ state, while each twofold defect $\sigma_e,\sigma_m,\sigma_\psi$ comes with two distinct species labels (e.g., $(\sigma_e)_{0,1}$), there is no non-trivial species label for threefold defects $\rho,\overline{\rho}$ as their symmetries mix even and odd QP sectors. We notice that in the cases when charge is fractionalized (i.e. the $A_r$ states for $r$ odd, and $D_r$ states for $r\equiv2,3$ mod 4), the $\mathbb{Z}_2$ species label also counts the fractional electric charge in units of $e^\ast/2$ carried by the defect (c.f. Ref.~\onlinecite{teo2013braiding}). %From this pattern we see that for a generic $ADE$ FQH state, in cases where there there are more distinct twist defect species, each defect contains less internal structure (quantum dimension). 
%the operation $1-\rho$ on the anyon lattice $\mathcal{A}_{so(8)}$ is equivalent to $\rho^{-1}$, which is invertible. And therefore $\mathcal{A}_{so(8)}/(1-\rho)\mathcal{A}_{so(8)}$ vanishes and there are no non-trivial species label for threefold defects. The operation $1-\sigma_\psi$ however is not invertible and has image $(1-\sigma_\psi)\mathcal{A}_{so(8)}=\mathbb{Z}_2=\{1,\psi\}$. And hence twofold defects have two distinct species.

\begin{table}[htbp]
\centering
\begin{tabular}{lll}
&defects and species&$d$\\\hline
$A_{2n}$&$\sigma=\sigma\times e$&$\sqrt{2n+1}$\\
$A_{2n+1}$&$\sigma_0=\sigma_0\times e^2$, $\sigma_1=\sigma_0\times e$&$\sqrt{n+1}$\\
$D_{2n}$&$\sigma_0=\sigma_0\times\psi$, $\sigma_1=\sigma_0\times \{e,m\}$ &$\sqrt{2}$\\
$D_{2n+1}$&$\sigma_0=\sigma_0\times e^2$, $\sigma_1=\sigma_0\times e$&$\sqrt{2}$\\
$D_4$&$\rho$, $\overline{\rho}$&2\\
&$(\sigma_\psi)_0=(\sigma_\psi)_0\times \psi$, $(\sigma_\psi)_1=(\sigma_\psi)_0\times e$&$\sqrt{2}$\\
&$(\sigma_e)_0=(\sigma_e)_0\times e$, $(\sigma_e)_1=(\sigma_e)_0\times m$&$\sqrt{2}$\\
&$(\sigma_m)_0=(\sigma_m)_0\times m$, $(\sigma_m)_1=(\sigma_m)_0\times\psi$&$\sqrt{2}$\\
$E_6$&$\sigma=\sigma\times e$&$\sqrt{3}$
\end{tabular}
\caption{Defects and species for the $A-D-E$ states, and their quantum dimensions $d$.}\label{tab:defectspecies}
\end{table}

%Species label $\lambda\in\mathbb{Z}_2$ for twofold defect can be statistically measured by a local Wilson measurement $\Theta_{\bf a}^\lambda$ of dragging a quasiparticle ${\bf a}$ twice around the defect. When compare to the bare defect $\mu_0$, $\Theta_{\bf a}^\lambda$ picks up an extra crossing phase \begin{align}\Theta_{\bf a}^\lambda/\Theta_{\bf a}^0=e^{2\pi i{\bf m}^TK^{-1}(1+\sigma){\bf a}}\label{lwm}\end{align} where ${\bf m}$ is the quasiparticle bound to the defect.
%Defect species $\lambda$ can be statistically measured by a local Wilson measurement $\Theta_{\bf a}^\lambda$ of dragging a quasiparticle ${\bf a}$ around the defect $d$ times, where $d$ is the order of the symmetry $M$ associated to the defect so that $M^d=1$. When compare to the bare defect $\mu_0$, $\Theta_{\bf a}^\lambda$ picks up an extra crossing phase \begin{align}\Theta_{\bf a}^\lambda/\Theta_{\bf a}^0=e^{2\pi i{\bf m}^TK^{-1}(1+M+\ldots+M^{d-1}){\bf a}}\label{lwm}\end{align} Eq.\eqref{lwm} is invariant under the change of equivalent defect bound quasiparticle ${\bf m}\to{\bf m}'={\bf m}+(M-1){\bf n}$.

Defect species can also undergo a {\em mutation} by absorbing or releasing a quasiparticle. This can be microscopically controlled, for example, by adding phase parameters in the sine-Gordon terms in \eqref{edgecoupling}, and letting them wind adiabatically by multiples of $2\pi$. The mutation process is summarized by the defect-quasiparticle fusion \begin{align}\sigma_{\lambda+p}=\sigma_\lambda\times e^p\end{align} where $\lambda\in\mathbb{Z}_2=\{0,1\}$ for the $A_{2n+1}$ and $D_{2n+1}$ states. For the $D_{2n}$ state, \begin{align}\sigma_{\lambda+1}=\sigma_\lambda\times e=\sigma_\lambda\times m,\quad\sigma_\lambda=\sigma_\lambda\times\psi\label{sigma1}.\end{align} Or for the $D_4$ state, $\sigma_\psi$ obeys \eqref{sigma1} and the other two twofold defects $\sigma_e$ and $\sigma_m$ follow \eqref{sigma1} up to cyclic permutation of quasiparticles.

The anti-partner of a defect $\mu$ with symmetry $M$ is a defect $\overline{\mu}$ with symmetry $M^{-1}$ and  reciprocal species label $\overline{\lambda}=-\lambda$. As $MM^{-1}=1$, a QP will not change type when dragged around the $\mu\times\overline{\mu}$ pair. This gives a Wilson measurement of the overall QP type associated with the defect \emph{pair}, known as an Abelian fusion channel. These channels are restricted only by defect species since QP parity (or half $e^\ast$ charge) is a conserved property. Fusions of twofold defect pairs in the $ADE$ states are
summarized in Table~\ref{tab:defectfusion}. They fix the quantum dimensions of defects (shown in Table~\ref{tab:defectspecies}) by identifying the total dimensions on both sides of the fusion equations. We see that fusion of two defects can give rise to a large number of types of QPs, and these possible internal states of two defects are precisely the same structure that allows them to compensate for the QP attachment emission-re-absorption constraint discussed above.

\begin{table}[htbp]
\centering
\begin{tabular}{ll}
&Fusion rules\\\hline
$A_{2n}$&$\sigma\times\sigma=1+e+\ldots+e^{2n}$\\
$A_{2n+1}$&$\sigma_0\times\sigma_0=\sigma_1\times\sigma_1=1+e^2+\ldots+e^{2n}$\\
&$\sigma_0\times\sigma_1=e+e^3+\ldots+e^{2n+1}$\\
$D_{2n}$&$\sigma_0\times\sigma_0=\sigma_1\times\sigma_1=1+\psi$\\
&$\sigma_0\times\sigma_1=e+m$\\
$D_{2n+1}$&$\sigma_0\times\sigma_0=\sigma_1\times\sigma_1=1+e^2$\\
&$\sigma_0\times\sigma_1=e+e^3$\\
$E_6$&$\sigma\times\sigma=1+e+e^2$
\end{tabular}
\caption{Defect pair fusion in the $A-D-E$ states.}\label{tab:defectfusion}
\end{table}

There is more structure in the $D_4=so(8)$ state due to the $S_3$ triality symmetry. $(\sigma_\psi)_\lambda\times(\sigma_\psi)_{\lambda'}$ obeys the same fusion rules as the $D_{2n}$ states in Table~\ref{tab:defectfusion}, while the two other twofold defects $\sigma_e,\sigma_m$ satisfy similar rules up to a cyclic permutation of quasiparticles. The threefold defect $\rho$ annihilates its anti-partner $\overline{\rho}$ and gives \begin{align}\rho\times\overline{\rho}=\overline{\rho}\times\rho=1+e+m+\psi.\label{eqn:sumrhorhobar}\end{align} This implies the quantum dimension $d_\rho=d_{\overline{\rho}}=2$ and the degenerate fusion of the pair \begin{align}\rho\times\rho=2\overline{\rho},\quad\overline{\rho}\times\overline{\rho}=2\rho.\label{eqn:sumrhorho}\end{align} The non-Abelian symmetry group $S_3$ results in non-commutative fusion rules \begin{align}\sigma_m\times\sigma_e&=\rho,\quad\sigma_e\times\sigma_m=\overline{\rho}\\\sigma_e\times\rho&=\rho\times\sigma_m=(\sigma_\psi)_0+(\sigma_\psi)_1.\end{align} All other fusion rules can be written down by 
cyclic 
permutation of the quasiparticle labels $e,m,\psi$.
 It is interesting to note here that  anyon condensation induced transitions in topological field theories studied in Ref.  \onlinecite{BaisSlingerlandCondensation} also point out the possibility for (confined) excitations with non-symmetric fusion rules. Though they do not give any explicit examples, it would be interesting to compare the underlying mechanisms in future work. 
 
\begin{figure}[htbp]
    \centering
    \includegraphics[width=0.4\textwidth]{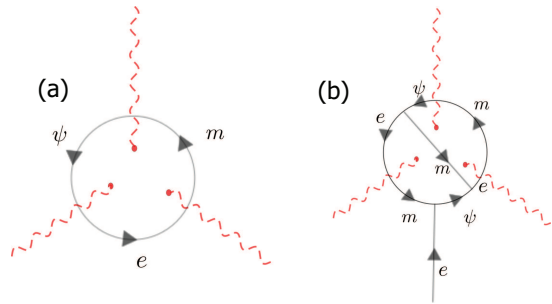}
    \caption{(a)Three $\rho$ branch cuts emerging from the center to create a vortex-like defect  (b) A quasiparticle string $e$  being absorbed by the defect configuration}
    \label{fig:three fold defect}
\end{figure}

 Another interesting property of the $D_4=so(8)$ is that the three-fold defects can be used to form a kind of twist vortex defect. This defect can be thought of as a point where (at least) three distinct gap edge phases meet at a point\cite{BarkeshliJianQi13, BarkeshliJianQi13long}, whereas, until now
 we have only considered point defects at the 
junction between two lines with different mass terms. 
We show an example of this ``vortex-like'' defect in
Fig. \ref{fig:three fold defect}a.
This defect is equivalent to the fusion of three $\rho$ defects. The fusion of three $\rho$ defects is also the same as considering the fusion of $\rho$ and $2\overline{\rho}$ from  \eqref{eqn:sumrhorho}.
Combining Eqs. \eqref{eqn:sumrhorho} and \eqref{eqn:sumrhorhobar} we find
\begin{equation}
\rho\times\rho\times\rho=\rho\times 2\overline{\rho}=2(1+e+m+\psi).
\label{eqn:rhorhorho}
\end{equation}
We can check this by matching the quantum dimensions: the quantum dimension of $\rho$ is 2 and the quantum dimensions of the Abelian particles on the right add up to the correct value. Furthermore, we see that the defect configuration
makes it possible to have a quasiparticle strings of all four types, i.e. $1,e,m,\psi,$  to be absorbed by the defect. In Fig. \ref{fig:three fold defect} we show an example
of an $e$ particle being absorbed at the defect site, similar configurations can also be drawn for $m,\psi$ by permuting the particle labels. This explains the 
existence of the different Abelian fusion channels.

%As explained earlier the fusion of a defect and its antidefect allows us to get a Wilson measurement of 
%the (in this case abelian) quasiparticle
%associated with the defect-antidefect pair. This is possible by Wilson loops labelled by quasiparticles.(the black circle in 
%Figure \ref{fig:three fold defect} for example).

\begin{figure}[H]
    \centering
    \includegraphics[width=0.25\textwidth]{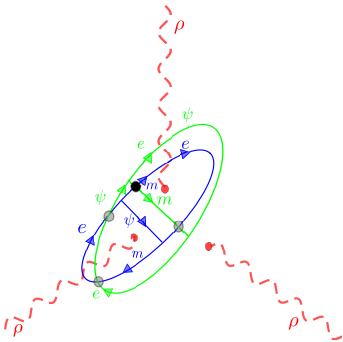}
    \caption{The quasiparticle labels for each loop are marked by their respective colors. The intersections which give a phase of $-1$ upon exchange
    are marked by grey circles. The intersection with a phase of $1$ is marked by a solid black circle.}
    \label{fig:doubling}
\end{figure}

We can also explain the factor of 2 in Eq. \ref{eqn:rhorhorho}  by drawing extra Wilson loops which form an  anti-commuting algebra, thus explaining the doubling of the Hilbert space.
This doubling  already happens at the fusion of two $\rho$ defects as seen in Eq. \ref{eqn:sumrhorho}. The extra structure already present for two $\rho$ defects allows us to draw additional 
Wilson loops which anticommute. This has been illustrated in Fig. \ref{fig:doubling}. The key point to note there is that the Wilson loops involve only two $\rho$ defects. 
Each of the blue and green loops exist independently and are separate Wilson lines. There are $4$ intersections between the blue and green loops, and since $2S_{e\psi}=2S_{em}=-1$ we pick up a phase of -1 three times (the grey dots)
and $S_{mm}=1$(the black dot) once. Thus we pick up a 
net phase of $-1$ when the two loops are exchanged. This explains the doubling of the Hilbert space.

% One loop that can be drawn is
% figure \ref{fig:problemloop}. However such loops commute with each other and with the loop in Figure \ref{fig:three fold defect}. Thus they cannot explain the factor of $2$ in equation \eqref{eqn:sumrhorhobar}.
% \begin{figure}[H]
%     \centering
%     \includegraphics[width=0.15\textwidth]{problemloop.eps}
%     \caption{}
%     \label{fig:problemloop}
% \end{figure}

\section{Discussion, Open Questions, and Conclusions}

 While we have not considered it here, we note that consideration of anyonic symmetries up to stable equivalence leads to an even
richer structure of  symmetries. So far we have 
considered anyonic symmetries satisfying Eq. \eqref{Autom}. These symmetries
preserve the $K$ matrix and the spins $h_{\mathbf{a}}$ of the quasiparticles,
$h_{\mathbf{a}}=\frac{1}{2}\mathbf{a}^TK^{-1}\mathbf{a}$. In other
words, the conformal structure on the edge is respected. However, we might relax this constraint and only require 
keeping the $T$ matrix invariant where
$T_{\mathbf{a},\mathbf{a}}=e^{2\pi i h_{\mathbf{a}}}$, then $h_{\mathbf{a}}$
will only be preserved mod 1. This will lead to more anyonic symmetries as
considered in Ref. \onlinecite{lu13}. For example, in the case of $su(12)$ the symmetries
which preserve spin lead to QP transformations 
generated by $e\rightarrow e^{-1}=e^{11}$ (c.f. Table \ref{tab:symmetries}) and we have
$h_e=h_{e^{-1}}=\frac{11}{24}$ (c.f. Table \ref{tab:Tmatrixtable}). However, if we only require keeping the $T$-matrix
invariant this leads to another anyonic symmetry generated by  $e\rightarrow
e^7$. In this case $h_{e^7}=1+\frac{11}{24}$. This is an
allowed symmetry because, $e^{2\pi i h_{e^7}}=e^{2\pi ih_e}$. Thus, if we require the full set of
anyonic symmetries which preserve the $T$ matrix, it will be generated by
\emph{both} the transformations $e\rightarrow e^7$ and $e\rightarrow e^{11}$. However demanding that
the symmetries preserve conformal structure on the edge restricts us to only
consider $e\rightarrow e^{11}$. 
We expect that these types of anyonic symmetries up to stable equivalence can
be realized by considering the
spin preserving symmetries of the the set of $K$ matrices $\{K':
K' \text{ is stably equivalent to K}\}.$ These will be the subject of future
work.

The above comments should not be taken to mean that our classification of defects and domain walls is incomplete or that our method is insufficient. As 
discussed in Appendix \ref{sec:stablequiv}, considering a matrix $K'$ which is stably equivalent to $K$ involves enlarging $K$ by some
$L\in GL(n,z)$ such that $K'=K\oplus L$ (possibly with some additional basis transformation $W$). $L$ describes a theory with equal numbers of right and left movers so that the chiral central charge is unchanged.
In these cases  the edge theory is \emph{no longer described by $K$ but by $K'$ }. The defects and gapped edges that can exist on the edge of $K'$ can again be analyzed 
using anyonic symmetry as defined in this article and will be comprehensive for that edge theory. We would like to emphasize that simply stating the anyonic symmetry in terms of the braiding matrix was not the goal of this article, but instead the focus has been on realizing distinct gapped edges and defects associated with that anyonic symmetry. Hence, once the edge theory has been \emph{fixed} by the $K$ matrix, we do not need to worry about extra anyonic symmetries that might arise from
stably equivalent matrices. Any gapped edge that can be written down for the theory must involve an expression of the form in equation 
\eqref{edgecoupling}, and all such terms can be fully classified by looking at the symmetries of the $K$ matrix alone.

 While we mentioned the connection between the fractional quantum Hall hierarchy states and the ADE series, the cases studied in this paper do not refer to states with a local fermionic 
 sector which is more relevant for experimentally observed filling fractions in two dimensional electron gas systems. The fermionic problem is more challenging.
For example, if $\mathbf{b}$ is a local boson and $\mathbf{a}$ an arbitrary anyon, then we identify $\mathbf{a}$ and $\mathbf{a+b}$
 and $\theta_{a}=\theta_{\mathbf{a+b}}=e^{2\pi i h_{\mathbf{a}}}=e^{2\pi i h_{\mathbf{a+b}}}.$ However, once we include fermions
 as local particles this can lead to complications such as $\theta_{\mathbf{a+f}}=-\theta_{a}$ because fermions have half integral spin.
 This leads to anyonic symmetries being $\mathbb{Z}_2$ graded according to fermion parity (which should be physically conserved), which adds more complexity but should be a straightforward, though perhaps technically challenging extension. 
 
To conclude, we associated the anyon relabeling symmetry of a general
Abelian topological phase with the group of outer automorphisms of the $K$-matrix.
%  {\color{blue} Although not shown explicitly in this article, we expect
% \eqref{outerdef} should depend only on the topological phase and be invariant
% under stable equivalence~\cite{cano2013bulk} of $K$-matrices, i.e.
% $\mbox{Outer}(K)=\mbox{Outer}(G^TKG)$ for unimodular $G$ and
% $\mbox{Outer}(K)=\mbox{Outer}(K\oplus\sigma_x)=\mbox{Outer}(K\oplus K_{E_8})$.}
We presented the AS of the bosonic $ADE$ Abelian topological states, and
discussed thoroughly the $\mathbb{Z}_2$ symmetry for $su(3)$ and $S_3$ triality
symmetry for $so(8)$. One dimensional gapped interface phases for chiral $ADE$
states were shown to be naturally classified by AS. A similar method was applied
to gapped edge phases for bosonic $ADE$ fractional quantum spin Hall states,
where extra constraints (in addition to the even and odd criterion in
Refs.~\onlinecite{LevinStern09, LevinStern12}) on the AS and 
the QP pair condensate were required for TR to be unbroken explicitly or spontaneously. It would be interesting to explore the interplay between AS and TR as well as the compatibility between QP pair condensates and TR in general fermionic phases. We studied topological point defects, each associated with an AS operation, and exhaustively described the fusion behavior of all possible twist defects in the $ADE$ states. Although not shown explicitly in this article, the $F$-symbols for defect state transformations should take a similar form to certain previously studied exactly solvable models in Refs.~\onlinecite{TeoRoyXiao13long, teo2013braiding}. These twist defects therefore form a consistent {\em fusion category}~\cite{Kitaev06}, and are powerful enough to construct a measurement-only topological quantum computer~\cite{BondersonFreedmanNayak08}.

{\bf{Note:}} During the preparation of the manuscript a recent work of Lu and Fidkowski appeared\cite{lu13}. Their nice work has similar themes and results to part of our work and both works compliment each other.  

{\emph{Acknowledgements}}
We acknowledge useful discussions with M. Barkeshli, L. Fidkowski, L. Kong, Y.-M. Lu, M. Mulligan, C. Nayak and X.-L. Qi. We thank J. Alicea for pointing us to Ref. \cite{lu13}. TLH was supported by ONR grant no. N0014-12-1-0935 (TLH). J.CY.T. acknowledges support from the Simons Foundation Fellowship. We also thank the support of the UIUC ICMT. 
\appendix
\section{Anyon theory of the bosonic Abelian $ADE$ states}
Here we summarize two dimensional bosonic Abelian quantum Hall states with $ADE$ chiral Kac-Moody (KM) current algebras at level one along boundary edges. We provide their $K$-matrices, quasiparticle (QP) labels, braiding, spin, and electric charge. %Throughout, we neglect the $E_8$ state which is topologically trivial.

The simplest $K$-matrices of the Chern-Simons actions that describe Abelian $ADE$ topological states are given by the Cartan matrices of the corresponding simply-laced algebras~\cite{KhanTeoHughesoappearsoon}. $A_r=su(r+1)$, for $r\geq2$, and $D_r=so(2r)$, for $r\geq4$, form infinite series of Abelian states, each with a $K$-matrix of rank $r$. \begin{align}
\left(K_{A_r}\right)_{IJ}&=2\delta_{IJ}-(\delta_{I,J+1}+\delta_{I,J-1})\\
\left(K_{D_r}\right)_{IJ}&=2\delta_{IJ}-(\delta_{I,J+1}+\delta_{I,J-1})+\nonumber\\
&\left(\delta_{I,r}\delta_{J,r-1}+\delta_{I,r-1}\delta_{J,r}-\delta_{I,r}\delta_{J,r-2}-\delta_{I,r-2}\delta_{J,r}\right).\end{align} There are also three exceptional simply-laced (i.e. with symmetric Cartan matrix) Lie algebras $K_{E_{r=6,7,8}}$ with $K$-matrices \begin{align}K_{E_6}=\left(
\begin{array}{cccccc}
 2 & -1 & 0 & 0 & 0 & 0 \\
 -1 & 2 & -1 & 0 & 0 & 0 \\
 0 & -1 & 2 & -1 & 0 & -1 \\
 0 & 0 & -1 & 2 & -1 & 0 \\
 0 & 0 & 0 & -1 & 2 & 0 \\
 0 & 0 & -1 & 0 & 0 & 2 \\
\end{array}
\right),\end{align}\begin{align} K_{E_7}=\left(
\begin{array}{ccccccc}
 2 & -1 & 0 & 0 & 0 & 0 & 0 \\
 -1 & 2 & -1 & 0 & 0 & 0 & 0 \\
 0 & -1 & 2 & -1 & 0 & 0 & -1 \\
 0 & 0 & -1 & 2 & -1 & 0 & 0 \\
 0 & 0 & 0 & -1 & 2 & -1 & 0 \\
 0 & 0 & 0 & 0 & -1 & 2 & 0 \\
 0 & 0 & -1 & 0 & 0 & 0 & 2
\end{array}
\right)\end{align}\begin{align}K_{E_8}=
\left(
\begin{array}{cccccccc}
 2 & -1 & 0 & 0 & 0 & 0 & 0 & 0 \\
 -1 & 2 & -1 & 0 & 0 & 0 & 0 & 0 \\
 0 & -1 & 2 & -1 & 0 & 0 & 0 & 0 \\
 0 & 0 & -1 & 2 & -1 & 0 & 0 & 0 \\
 0 & 0 & 0 & -1 & 2 & -1 & 0 & -1 \\
 0 & 0 & 0 & 0 & -1 & 2 & -1 & 0 \\
 0 & 0 & 0 & 0 & 0 & -1 & 2 & 0 \\
 0 & 0 & 0 & 0 & -1 & 0 & 0 & 2 \\
\end{array}
\right)\end{align}

\begin{figure}[ht]
\centering
\includegraphics[width=0.5\textwidth]{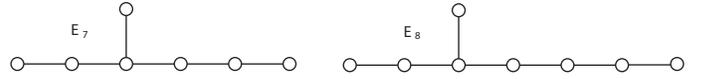}
\caption{Dynkin diagrams for the exceptional simply-laced Lie algebras $E_7,E_8$.}\label{fig:DynkinE7E8}
\end{figure}

Since all diagonal entries are $2$, any such state is bosonic as all local particles are bosons. The $E_8$ state does not have topological order as $\det(K_{E_8})=1$ so that the anyon content $\mathcal{A}_{E_8}=\mathbb{Z}^8/K_{E_8}\mathbb{Z}^8=1$ is trivial and all QPs are mutually local. The $K$-matrices (or Cartan matrices) can also be read off from the Dynkin diagrams of the Lie algebras (see Fig.~ 2 for $A_r,D_r,E_6$ and Fig.~\ref{fig:DynkinE7E8} for $E_7,E_8$). By assigning an enumeration $I=1,\ldots,r$ of the dots in the Dynkin diagram, the non-zero entries of the $K$-matrices are given by $K_{II}=2$ and $K_{IJ}=-1$ if dot $I$ and $J$ are connected. 

Even though we are considering purely bosonic states, it is important to note that each Abelian $ADE$ state can be realized in an electronic system exhibiting a fractional quantum Hall (FQH) state. The $K$-matrix of such an electronic $ADE$ state needs to be modified to include local fermionic electrons. For instance a $K$-matrix could take the form of a direct sum $\mathcal{K}=K_{ADE}\oplus\sigma_z$~\cite{FrohlichThiran94, FrohlichStuderThiran97}, where $\sigma_z$ only introduces local fermions. The electric charge vector would take the form ${\bf t}=(2,\ldots,2,1,1)$ so that primitive local bosons are treated as charge $e^\ast=2e$ pairs of electrons. As the additional $\sigma_z$ does not contribute to the topological order ($|\det(\sigma_z)|=1$), all results in the main text extends to such electronic FQH states.

Next we describe the anyon lattice vectors corresponding to the QPs in an Abelian $ADE$ topological state. The bulk boundary correspondence identifies a bulk QP to an edge vertex operator, $\psi^{\bf{a}}\sim e^{i{\bf a}\cdot\boldsymbol\phi}$, where $\bf{a}$ is an $r$-dimensional anyon lattice vector in $\Gamma^\ast=\mathbb{Z}^r$. It has charge $q_{\bf a}e^\ast$ for $q_{\bf a}={\bf t}^TK^{-1}{\bf a}$, where the charge vector ${\bf t}=(1,1,\cdots,1)$ describes the external electromagnetic coupling with fundamental charge $e^\ast$ for local bosons. As explained in the main text, due to the local boson condensate at zero temperature, the QPs are defined only up to local particles, i.e. ${\bf a}\equiv{\bf a}+K\mathbb{Z}^r$. And the fractional electric charge is defined modulo integer multiples of $e^\ast$. Table~\ref{tab:Quasiparticle_labels} lists a particular representation of each set of equivalent anyons $[{\bf a}]={\bf a}+K\mathbb{Z}^r$.  
\begin{table}[htbp]
\centering
\begin{tabular}{l|c|l}
Algebra& Anyon & Anyon
vector\\\hline
{$A_r\equiv su(r+1)$} & $1$
&$(0,\cdots,0)$\\
$r\geq 2$&$e^i$ ($1\leq i\leq r$)&$(0,\cdots,\underbrace{1}_{i^{th}},\cdots,0)$\\\hline
$D_r$($r$ odd)&1&$(0,\cdots,0)$\\
$r>4$&$e$&$(0,\cdots,0,1,0)$\\
&$e^2$&$(1,\cdots,0)$\\
&$e^3$&$(0,\cdots,0,1)$\\\hline
$D_r$($r$ even)&1&$(0,\cdots,0)$\\
$r>4$&$e$&$(0,\cdots,0,1,0)$\\
&$m$&$(0,\cdots,0,1)$\\
&$\psi$&$(1,\cdots,0)$\\\hline
$E_6$&$1$&$(0,0,0,0,0,0)$\\
&$e$&$(1,0,0,0,0,0)$\\
&$e^2$&$(0,0,0,0,1,0)$\\\hline
$E_7$&$1$&$(0,0,0,0,0,0,0)$\\
&$e$&$(0,0,0,0,0,1,0)$\\
\end{tabular}
\caption{$r$-dimensional quasiparticle vectors of the $ADE$ Abelian topological states at level 1.}\label{tab:Quasiparticle_labels}
\end{table}
\begin{table}[htbp]
\centering
\begin{tabular}{l|l}
Algebra & Fractional electric charge
\\\hline
{$A_r$}($r$ even) & All anyons are neutral\\
&(or have integral charge)\\\hline
{$A_r$}($r$ odd) & $e^i$ $i$ even$\rightarrow$ integral charge\\
& $e^i$ $i$ odd$\rightarrow$ half-integral charge
\\\hline
{$D_r$}($r$ mod $4=$ $0,1$) & All anyons are neutral\\
&(or have integral charge)\\\hline
{$D_r$}($r$ mod $4=$ 2) & $1,\psi\rightarrow$ neutral (or integral charge)\\
& $e,m\rightarrow$ half-integral charge
\\\hline
{$D_r$}($r$ mod $4=$ 3) & $1,e^2$ $\rightarrow$ neutral (or integral charge)\\
& $e,e^3\rightarrow$ half-integral charge\\\hline
$E_6$&All anyons are neutral\\
&(or have integral charge)
\\\hline
$E_7$& $1\rightarrow$ neutral\\
&$e\rightarrow$ half-integral charge
\\\hline
\end{tabular}
\caption{Fractional electric charges of the quasiparticles in units of fundamental boson charge $e^\ast$.}\label{tab:Quasiparticle_charges}
\end{table}

QP braiding is summarized by the $S$ matrix, where the phase of \begin{align}S_{{\bf a}{\bf b}}=\frac{1}{\mathcal{D}}e^{2\pi i{\bf a}^TK^{-1}{\bf b}}\end{align} corresponds the braiding phase if the QP $\psi^{\bf a}$ is dragged once around $\psi^{\bf b}$. The normalization $\mathcal{D}=\sqrt{\det(K)}$, known as the total quantum dimension, is to ensure unitarity of the $S$ matrix. The topological spin of a QP is given by the exchange phase \begin{align}\theta_{\bf a}=e^{2\pi ih_{\bf a}}=e^{\pi i{\bf a}^TK^{-1}{\bf a}}.\end{align} The spin of the quasiparticle is 
\begin{align}
h_{\bf{a}}&=\frac{1}{2}\mathbf{a}^tK^{-1}\mathbf{a}.
\end{align}
The braiding phase and exchange phase (topological spin) are invariant under the addition of local bosons. 
The QP's spin and braiding phases are listed in Table~\ref{tab:Smatrixtable}, and \ref{tab:Tmatrixtable} respectively, and are labeled according to the anyon labels in Table~\ref{tab:Quasiparticle_labels}. For instance they verify the triality $S_3=Dih_3$ symmetry for $so(8)$ and the eightfold periodicity $D_r\rightarrow D_{r+8}$.
\begin{table}[t]
\centering
\begin{tabular}{l|l}
Algebra & $S$ matrix
\\\hline
{$A_r$} & $S_{e^\mu,e^\nu}=\dfrac{1}{\sqrt{r+1}}\exp\left[-2\pi i
\frac{\mu\nu}{r+1}\right]$\\
&$0\leq \mu,\nu\leq r$
\\\hline
{$D_r$}($r$ mod $4=$ $0$) & $S_{ee}=S_{mm}=S_{\psi\psi}=1$\\
&$S_{em}=S_{e\psi}=S_{m\psi}=-1$\\\hline
{$D_r$}($r$ mod $4=$ $1$) &  $S_{e^\mu,e^\nu}=\dfrac{1}{2}\exp\left[\pi i
\frac{\mu\nu}{2}\right]$ $0\leq \mu,\nu\leq 3$\\\hline
{$D_r$}($r$ mod $4=$ $2$) & $S_{ee}=S_{mm}=S_{e\psi}=S_{m\psi}=-1$\\
&$S_{em}=S_{\psi,\psi}=1$\\\hline
{$D_r$}($r$ mod $4=$ $3$) & $S_{e^\mu,e^\nu}=\dfrac{1}{2}\exp\left[-\pi i
\frac{\mu\nu}{2}\right]$ $0\leq \mu.\nu\leq 3$ \\\hline
$E_6$& $S_{e,e}=S_{e^2,e^2}=S^*_{e,e^2}=\frac{1}{\sqrt{3}}\exp[\frac{2\pi
i}{3}]$
\\\hline
$E_7$& $S_{e,e}=-\frac{1}{\sqrt{2}}$
\\\hline
\end{tabular}
\caption{Braiding $S$-Matrix of anyons}\label{tab:Smatrixtable}
\end{table}
\begin{table}[t]
\centering
\begin{tabular}{l|l}
Algebra & Spin
\\\hline
{$A_r$} & $h_{\mu}=\frac{\mu}{2}(1-\frac{\mu}{r+1})$\\
&$0\leq \mu\leq r$
\\\hline
{$D_r$}($r$ odd) & $h_1=0;h_e=h_{e^3}=\frac{r}{8};h_{e^2}=\frac{1}{2}$\\\hline
{$D_r$}($r$ even) &  $h_1=0;h_e=h_m=\frac{r}{8};h_{\psi}=\frac{1}{2}$\\\hline
$E_6$& $h_1=0;h_e=h_{e^2}=\frac{2}{3}$
\\\hline
$E_7$& $h_1=0;h_e=\frac{3}{4}$
\\\hline
\end{tabular}
\caption{Quasiparticle Spin.}\label{tab:Tmatrixtable}
\end{table}

%%%%%%%%%%%%%    Outer(K) and Inner(K) for su(3)           %%%%%%%%%%%%%%%%%%%
\section{Anyonic relabeling symmetry $Outer(K)$ of $su(3)$}\label{sec:autsu3}
In the main text, we defined the notion of {\em anyon relabeling symmetry} by the group of outer automorphisms in
Eq. 7. Here we demonstrate this explicitly for the $A_2=su(3)$ state. 
The group of automorphisms $\mbox{Aut}(K)$ can be identified with the dihedral group $Dih_6$, 
the symmetry group of a hexagon. It is is generated by a sixfold ``rotation" $P$ and a twofold ``reflection" $R$, 
and has the two-dimensional representation

\begin{widetext}
\begin{align}\mbox{Aut}\left(K_{su(3)}\right)=\left\langle\left. P=\left(
\begin{array}{cc}
 0 & -1 \\
 1 & 1 \\
\end{array}
\right),R=\left(
\begin{array}{cc}
 0 & 1 \\
 1 & 0 \\
\end{array}
\right)\right|P^6=R^2=1,RPR=P^{-1}\right\rangle .\end{align}
\end{widetext}
We notice that these matrices are isometries with respect to the $K$-matrix for $su(3)$, $PKP^T=RKR^T=K$. 
Also, $P$ and $R$ act on anyon labels by taking $e\leftrightarrow e^2$. $P$ and $R$ can be 
visualized  geometrically in Figs. 1(c) and 1(d) respectively.
On the other hand the matrices $PR,P^2$  preserve anyon labels up to local particles, 
and therefore generate the group of inner automorphisms.
$P^2$ is geometrically represented in Fig. 1(a) and Fig. 1(b) represents the effect of $RP^3$. 
\begin{align}\mbox{Inner}\left(K_{su(3)}\right)=\left\langle PR,P^2\right\rangle=\mathbb{Z}_2\ltimes\mathbb{Z}_3=S_3.\end{align} The quotient \begin{align}\mbox{Outer}\left(K_{su(3)}\right)=\frac{\text{Aut}(K)}{\text{Inner}(K)}=\frac{Dih_6}{S_3}=\mathbb{Z}_2=\{1,\sigma\}\end{align} describes a mirror symmetry of the $A_2=su(3)$ state, $e\leftrightarrow e^2$. We further notice that the equivalence class $1$ can be represented by any of the elements $1,P^2,P^4,PR,P^3R,P^5R$, while the equivalence class $\sigma$ can be represented by $R,P,P^3,P^5,P^2R,P^4R$. However, if we further impose the constraint of charge conservation, we must leave ${\bf{t}}={(1,1)}$ invariant, and thus we are 
restricted 
to the identity matrix for conjugacy class $1$ and $R$ for mirror $\sigma$.

%%%%%%  General charge preserving symmetry matrix table%%%%%%%%%%%%%%%%%%%%%%%%%%%%%%%%
For a general $ADE$ state, $\mbox{Outer}(K)=\mathbb{Z}_2$ except for $D_4=so(8)$, where $\mbox{Outer}(K_{so(8)})=S_3$ as explained in the main text. Given any $ADE$ state, it is always possible to find a \emph{charge conserving} symmetry matrix which realizes the symmetry. It has been already explicitly written down in the main text for the $A_2=su(3)$ and $D_4=so(8)$ states. The charge conserving matrices that represent the mirror anyonic symmetries $\sigma$ in Table~II  for other simply-laced Lie algebras are listed in Eq.~\eqref{sigmaADE}. There is no mirror anyonic symmetry for $E_7$ as its symmetry group Outer(K) is trivial.

\begin{align}\sigma_{A_r}=\left(\begin{array}{*{20}c}0&\ldots&0&1\\0&\ldots&1&0\\\vdots&\ddots&\vdots&\vdots\\1&\ldots&0&0\end{array}\right)_{r\times r}&\sigma_{D_r}=\left(\begin{array}{*{20}c}\openone&0\\0&\sigma_x\end{array}\right)_{r\times r},\nonumber\\\sigma_{E_6}=&\left(\begin{array}{cccccc}
 0 & 0 & 0 & 0 & 1 & 0 \\
 0 & 0 & 0 & 1 & 0 & 0 \\
 0 & 0 & 1 & 0 & 0 & 0 \\
 0 & 1 & 0 & 0 & 0 & 0 \\
 1 & 0 & 0 & 0 & 0 & 0 \\
 0 & 0 & 0 & 0 & 0 & 1 \\
\end{array}\right).\label{sigmaADE}
\end{align}

\section{Outer Automorphisms of Lie Algebras and anyonic symmetry}\label{sec:Outer Automorphisms}
We provide some further details and references to elucidate the structure of the outer automorphisms of a Lie Algebra and
its relation to anyonic symmetry. This section is rather mathematical and is not necessary to understand the rest of the paper.

Given a Lie algebra we have a set of generators $\mathrm{g}$. An automorphism of a Lie algebra is a map $\omega:\mathrm{g}\rightarrow\mathrm{g}$ which
is linear and respects the structure of the Lie algebra, i.e. $\omega([x,y])=[\omega(x),\omega(y)],x,y\in\mathrm{g}$. The set of automorphisms $\text{Aut}(\mathrm{g})$
form a group called the automorphism group of the Lie algebra. It has a normal subgroup, denoted by $\text{Inn}(\mathrm{g})$ which is generated by
\begin{eqnarray*}
 &\exp{\left(\text{ad}_x\right)}:x\in\mathrm{g};\nonumber\\&\text{where}\quad ad_x:\mathrm{g}\rightarrow\mathrm{g}, \text{ad}_x(y)=[x,y],\,x,y\in\mathrm{g}
\end{eqnarray*}
and is called the set of inner automorphisms.

Since $\text{Inn}(\mathrm{g})$ is a normal subgroup of $\text{Aut}(\mathrm{g})$ we can define the coset
\begin{align*}
 \text{Out}(\mathrm{g})=\frac{\text{Aut}(\mathrm{g})}{\text{Inn}(\mathrm{g})},
\end{align*}
which is the group of outer automorphisms. It is a well known theorem in mathematics that $\text{Out}(\mathrm{g})$ is isomorphic to the automorphisms of the Dynkin diagram.
The interested reader can look up the proofs in proposition $D.40$ of Ref. \onlinecite{fulton1991representation}, Ch 11 of Ref. \onlinecite{fuchs1995affine} or 
section 12.2, in particular Table 1 of Ref. \onlinecite{humphreys1972introduction}.This connection has been exploited in the main text to deduce the outer automorphisms
of the various Lie algebras and, as we discuss below, has motivated our definition of anyonic symmetries.

The discussion henceforth will be limited to simply laced Lie algebras with symmetric Cartan matrices. All the information about a Lie
algebra is encoded in the Cartan matrix and the associated Dynkin diagram (c.f. Figs. \ref{fig:Dynkin},\ref{fig:DynkinE7E8}) of the Lie algebra. The number of nodes of the Dynkin
diagram is equal to the rank of the Lie algebra.

Let us consider a simply laced Lie algebra with rank $r$. In this case the weight lattice (the anyon lattice) $\Gamma^{\ast}=\mathbb{Z}^r$ and the root
lattice $\Gamma=K\mathbb{Z}^r$ where $K$ is the Cartan matrix. The Cartan matrix induces an inner product in the weight space
\begin{align*}
 \langle\mathbf{a},\mathbf{b}\rangle=\mathbf{a}^TK^{-1}\mathbf{b}\quad\mathbf{a,b}\in\mathbb{Z}^r.
\end{align*}
A basis for the root lattice of the Lie algebra (i.e., the simple roots) is given by the vectors
\begin{align*}
 \alpha_i=(K_{i1},K_{i2},\cdots,K_{ir}),\quad i\in[1,\cdots,r].
\end{align*}
In mathematics the components of $\alpha_i$ are known as the Dynkin labels.
In the normalization $\alpha_i^2=2$ the Cartan matrix is $K_{ij}=\langle\alpha_i,\alpha_j\rangle$.
Thus, if $\omega\in \text{Aut}(\mathrm{g})$, $\langle\alpha_i,\alpha_j\rangle=\langle\omega\alpha_i,\omega\alpha_j\rangle=K_{ij}$

Now we are in a position to motivate the definition of $\text{Aut}(K)$ in equation \eqref{Autom}. The automorphisms 
induce a linear transformation in the space of roots such that the Cartan matrix is preserved. The transformation $M$ can be interpreted
simply as a basis change for the Cartan matrix. But since the Cartan matrix stays invariant, we have defined the set of automorphisms 
$\mbox{Aut}(K)=\left\{M\in GL(r;\mathbb{Z}):MKM^T=K\right\}\label{Aut(K)}.$ The fact that $M\in GL(r;\mathbb{Z})$ just implies that this is 
a volume preserving basis transformation which preserves the number of ground states $\text{det}(K)$. The roots transforming among themselves
indicate that local particles stay local.

Before we carry on, we note that the the automorphisms preserve the Dynkin diagrams and the Cartan matrix. However, they induce rotations
in the space of weights $\mathbb{Z}^r$. The anyons live in the \emph{weight} space. Each node of the Dynkin diagram can be associated with a corresponding
fundamental weight. The $i$-th node will have Dynkin label $(0,\cdots,1,\cdots,0)$ with $1$ in the $i$-th entry. Thus inner automorphisms act trivially in the
weight space, thus preserving the weight labels(up to roots) whereas outer automorphisms act non trivially 
on the weight space rotating distinct weights (anyons) into each other.
This can help one understand the definition of inner automorphism in the context of anyons.
 \begin{align*}\mbox{Inner}(K)=\left\{M_0\in\mbox{Aut(K)}:[M_0{\bf a}]=[{\bf a}]\right\}.\end{align*}
Since we are only interested in transformations which
interchange weights we quotient out $\text{Inn}(K)$ to get $\frac{\text{Aut}(K)}{\text{Inn}(K)}$ as the group of anyon symmetries.

\subsection{An example $su(3)_1$}
Now let us consider an example in particular: $su(3)_1$. We have already seen explicit realization of the automorphism group of $su(3)$ in Appendix
\ref{sec:autsu3}. We now see the action of inner an outer automorphisms on the roots and weights of $su(3)$. 

The Cartan matrix of $su(3)$ is 
$K=\begin{pmatrix}
2&-1\\
-1&2
\end{pmatrix}
$.

A basis set for the root system is given by $\alpha_1=(2,-1)$ and $\alpha_2=(-1,2)$. Hence, the roots obey
\begin{align}
 \langle\alpha_1,\alpha_1\rangle=\langle\alpha_2,\alpha_2\rangle=2;\,\, \langle\alpha_1,\alpha_2\rangle=-1.
 \label{eq:su3innprod}
\end{align}
These roots correspond to local particles. On the other hand the weights correspond to the anyon labels
$e_1=(1,0)$ and $e_2=(0,1)$. Now we explicitly examine the action of the inner and aouter automorphisms on the roots and the weights
to reinforce the statements made above.

As shown in appendix \ref{sec:autsu3}, inner automorphisms are generated by $PR,P^2$ where $P=\left(
\begin{array}{cc}
 0 & -1 \\
 1 & 1 \\
\end{array}
\right),R=\left(
\begin{array}{cc}
 0 & 1 \\
 1 & 0 \\
\end{array}
\right)$.
\begin{itemize}
 \item{Inner Automorphisms}\newline
 The action of inner automorphisms are 
 \begin{align}
  PR\alpha_1&=-\alpha_ 1\nonumber\\
  PR\alpha_2&=\alpha_1+\alpha_2\nonumber\\
  \text{Using equation \ref{eq:su3innprod} }& \langle\alpha_i,\alpha_j\rangle=\langle PR\alpha_i,PR\alpha_j\rangle,\,[i,j]\in[1,2]\nonumber\\
  PRe_1&=e_1-\alpha_1\nonumber\\
  PRe_2&=e_2
 \end{align}
\begin{align}
   P^2\alpha_1&=\alpha_2\nonumber\\
  P^2\alpha_2&=-\alpha_1-\alpha_2\\
 \langle\alpha_i,\alpha_j\rangle&=\langle P^2\alpha_i,P^2\alpha_j\rangle,\,[i,j]\in[1,2]\nonumber\\
 P^2e_1&=e_1-\alpha_1\equiv e_1\nonumber\\
  P^2e_2&=e_2-\alpha_1-\alpha_2\equiv e_2.
\end{align}
Thus we see that the inner automorphisms induce linear maps on the set of roots which preserve the Cartan matrix.
They act trivially on the weight space too, thus the weights (anyon labels) are preserved up to local particles.
Hence the inner automorphisms act trivially on the weight space(up to roots).

\item{Outer Automorphisms}\\
The outer automorphisms act by
 \begin{align}
  P\alpha_1&=\alpha_ 1+\alpha_ 2\nonumber\\
  P\alpha_2&=-\alpha_1\nonumber\\
 \langle\alpha_i,\alpha_j\rangle&=\langle P\alpha_i,P\alpha_j\rangle,\,[i,j]\in[1,2]\nonumber\\
  Pe_1&=e_2\nonumber\\
  Pe_2&=e_1-\alpha_1
 \end{align}
\begin{align}
   R\alpha_1&=\alpha_2\nonumber\\
  R\alpha_2&=\alpha_1\nonumber\\
 \langle\alpha_i,\alpha_j\rangle&=\langle R\alpha_i,R\alpha_j\rangle,\,[i,j]\in[1,2]\nonumber\\
 Re_1&=e_2\nonumber\\
  Re_2&=e_1.
\end{align}
\end{itemize}
Hence we see that while the linear transformations induced by the outer automorphisms preserve the Cartan matrix they act nontrivially
on the weight space (anyon labels) interchanging them.

\section{Stable equivalence}\label{sec:stablequiv}
We introduce the basic definitions of stable equivalence in this appendix.
More detailed expositions and stronger results can be found in Refs. \onlinecite{cano2013bulk,BarkeshliJianQi13long}.

Two $K$ matrices $K_1$ and $K_2$ of the same dimension($N$) and signature are 
stably equivalent if there exist signature $(n,n)$ unimodular matrices $L_1$ and $L_2$ such that there exists $W\in GL(N+2n,\mathbb{Z})$ so that
\begin{align}
K_1\oplus L_1&=W^{T}\left(K_2\oplus L_2\right){W}.
 \label{eq:stableequivalencedefinition}
\end{align}
The signature of a matrix is determined by $n_+-n_-$ where $n_+$ and $n_-$ are the number of positive and negative eigenvalues. Physically this 
is the chiral central charge of the edge theory described by the $K$ matrix. In general if the $S$ and $T$ matrices characterizing the anyon content 
of two $K$ matrices are the same, then they are stably equivalent\cite{cano2013bulk}. Physically, stable equivalence 
was introduced in the context of edge reconstruction of a quantum Hall state by 
adding trivial degrees of freedom to the edge and can lead to the interesting situation where multiple edge state theories can support the same bulk topological phase. If we restrict ourselves to bosonic quantum Hall states
(as we do in this article), $L_1$ and $L_2$ must be even (they must have even entries on the diagonal).

\section{Stability of gapped interface phase}
The set of sine-Gordon coupling terms in Eq. 8 introduces a finite
energy gap for all of the edge degrees of freedom, and correspond to a particular gapped phase along the interface. By
rearranging the bosons $\phi^R_I=\varphi+(K^{-1})_{IJ}\theta_J$ and
$\phi^L_I=\varphi_I-(K^{-1})_{IJ}\theta_J$, Eq. 8 becomes an
ordinary gapped sine-Gordon model
\begin{align}\mathcal{L}_M=\frac{1}{\pi}
\partial_x\varphi_I\partial_t\theta_I-g_I\cos(2\tilde\theta)\end{align} where
$\langle2\tilde\theta_I\rangle=\langle
K_{IJ}(\phi^R_J+M_{J'J}\phi^L_{J'})\rangle$ is pinned at $0$ modulo $2\pi$ when
$g_I$ is large for all $I.$ %The $\varphi$ degree of freedom can be integrated out through
%incorporating a boson Hamiltonian density
%\begin{align}\mathcal{H}=V_{IJ}^{\sigma\sigma'}
%\partial_x\phi_I^\sigma\partial_x\phi^{\sigma'}_J\end{align}
%\begin{align}\mathcal{H}=V_{IJ}^{\varphi\varphi}
%\partial_x\varphi_I\partial_x\varphi_J+V_{IJ}^{\theta\theta}
%\partial_x\theta_I\partial_x\theta_J+V_{IJ}^{\varphi\theta}
%\partial_x\varphi_I\partial_x\theta_J\end{align}

In the strong coupling limit, the collection of backscattering terms
$\{g^{(M)}_I\}$ with respect to a symmetry $M$ in Eq. 8 
describes a fully gapped interface phase $\mathcal{L}_M$. Their associated scaling dimensions
$\Delta(g^{(M)}_I)$ determine the low energy relevance at the fixed point in the
renormalization group sense. Similar to conventional Luttinger liquid theory, they depend
on the forward scattering Hamiltonian
\begin{align}\mathcal{H}=V_{IJ}^{\sigma\sigma'}
\partial_x\phi_I^\sigma\partial_x\phi^{\sigma'}_J.\end{align} The backscattering
terms $g^{(M)}_I$ can be simultaneously tuned to be relevant by an appropriate
choice of $V^{\sigma\sigma'}_{IJ}$ ~\cite{MooreWenedgerelevance1,
MooreWenedgerelevance2, cano2013bulk}.
\section{Time Reversal Invariance(TRI) for bosonic topological theories (bulk and edge) }
\subsection{General formulation}
In this subsection we begin with a brief discussion of TRI for bosonic systems, without any reference to the particular form 
of the $K$ matrix or charge vector $\mathbf{t}$. We follow the discussions in Refs. 
\onlinecite{LevinStern12,LuVishwanathE8}.

Let us consider the bulk CS Lagrangian
\begin{align}
  \mathcal{L}_{\text{bulk}}&=\frac{K_{IJ}}{4\pi}\epsilon^{\mu\nu\lambda}\alpha_{I\mu}\partial_{\nu}\alpha
	_{J\lambda}-\frac{e^{\ast}}{2\pi}t_{I}\epsilon^{\lambda\mu\nu}A_{\lambda}\partial_{\mu}\alpha_{I\nu},\nonumber\\
	&\quad I,J=1,2,\cdots,N.
\label{eq:bulkCSconv}
\end{align}
We impose TRI on the system and study the implications on the bulk and the edge.
Here we have assumed that there are $N$ $U(1)$ gauge fields $\alpha_I$, thus the $K$ matrix is $N\times N$ and the gauge 
group is $U(1)^N$.
The system being bosonic, the diagonal entries of the $K$ matrix must be even
\begin{align*}
	K_{II}=0\;\; \text{mod}\; 2.
\end{align*}

Under the action of the anti-unitary time reversal operator
$\mathcal{T}$, the external electromagnetic gauge field $A$ transforms as 
\begin{align*}
  A^0&\rightarrow A^0\\
A^i&\rightarrow -A^i;\,\, i=1,2.
\end{align*}
$\mathcal{T}$ acts on the internal CS gauge fields $\alpha_I$ as
 \begin{align}
   \alpha_{I\mu}\rightarrow \mp T_{IJ}\alpha_{J\mu}   
   \label{eq:TimereversalCS}
 \end{align}
 where the $-$ sign corresponds to the
time index $\mu=0$ and the $+$ sign corresponds to the spatial indices $\mu=1,2$.
 Here,
$T$ is an integer valued $N\times N$ matrix which has to obey some constraints as outlined below. When we impose TRI on
$\mathcal{L}_\text{bulk},$  and using equation \eqref{eq:TimereversalCS}, we find the requirement
\begin{align}
T^TKT&=-K\label{eq:TimereversalK1}\\
T\mathbf{t}=\mathbf{t}.
\label{eq:TimereversalK}
\end{align}
For a bosonic system, $\mathcal{T}^2=\mathbf{1}$. Hence,
\begin{align}
	T^2=\mathbf{1}.
	\label{eq:tsquared}
\end{align}
Next, let us consider the edge of the system in \eqref{eq:bulkCSconv}. 
The Lagrangian density is
\begin{align}
  \mathcal{L}_{\text{edge}}&=\frac{K_{IJ}}{4\pi}\partial_x\phi_I\partial
	_t\phi_J+\frac{e^{\ast}}{2\pi}\epsilon^{\mu\nu}t_I\partial_{\mu}\phi_IA_\nu+\nonumber\\& \text{forward
	scattering terms}.
\label{eq:edgeCSconv}
\end{align}
Remembering that $\partial_{\mu}\boldsymbol{\phi}=\boldsymbol{\alpha}_{\mu}$\cite{Wenbook,Wenedgereview} and using equation \eqref{eq:TimereversalCS}, we get
\begin{align*}
  {\mathcal{T}}^{-1}\phi_I\mathcal{T}&=T_{IJ}\phi_J+C_I, C_I\in \mathbb{R}.
\end{align*}
Here $C_I$ is a constant which will be fixed later from physical considerations.
For notational convenience and to align our expressions with previous work, let us replace $C_I$ by $\pi\left(  K^{-1} \right)_{IJ}\chi_J,$ i.e.,
\begin{align}
  \mathcal{T}^{-1}\phi_I\mathcal{T}&=T_{IJ}\phi_J+\pi\left(  K^{-1}\right)_{IJ}\chi_J,\,\,\boldsymbol{\chi}\in\mathbb{R}^N.
	\label{eq:edgetimereversal}
\end{align}
$\boldsymbol{\chi}$ is often referred to as the time reversal vector. 

Physically $\boldsymbol{\chi}$ determines the action of time reversal $\mathcal{T}$ on vertex operators 
$\psi^{\mathbf{a}}=e^{i\mathbf{a}\cdot\boldsymbol{\phi}}$ on the edge. However,  different $\boldsymbol{\chi}$'s  are not necessarily physically distinct.
In fact, they might be gauge equivalent to each other.
In order to understand this we start off by noting that  $\mathcal{L}_\text{edge}$ is
shift invariant in $\boldsymbol{\phi}$. Thus, $\boldsymbol{\phi}\rightarrow\boldsymbol{\phi}+\boldsymbol{\xi}$
leaves $\mathcal{L}_\text{edge}$ unchanged. This should come as no surprise 
as the gauge field $\alpha_{I\mu}$ is also left unchanged by this redefinition. Indeed,
$\alpha_{I\mu}=\partial_{\mu}(\phi_I+\xi_I)=\partial_{\mu}\phi_{I}.$ But, $\boldsymbol{\phi}\rightarrow\boldsymbol{\phi}+\boldsymbol{\xi}\implies
e^{i\mathbf{a}\cdot\boldsymbol{\phi}}\rightarrow e^{i\mathbf{a}\cdot\boldsymbol{\phi}}e^{i\mathbf{a}\cdot\boldsymbol{\xi}}$.\\
Now we can consider the {\emph{global}} $U(1)$ gauge transformation on the edge associated with the shift invariance of $\boldsymbol{\phi}$, 
\begin{equation}
  \tilde{\psi}^{\mathbf{a}}=\psi^{\mathbf{a}}e^{i\mathbf{a}\cdot\boldsymbol{\xi}}\quad\text{where }\psi^{\mathbf{a}}=e^{i\mathbf{a}\cdot\boldsymbol{\phi}}.
  \label{eq:gaugetransformation}
\end{equation}
With this we can see how $ \tilde{\psi}^{\mathbf{a}}$ and $\psi^{\mathbf{a}}$ transform under TR:
\begin{align}
  \mathcal{T}^{-1}\psi^{\mathbf{a}}\mathcal{T} &=\left[ \psi^{\left( T^T\mathbf{a} \right)} \right]^\dagger e^{-i\pi\left( K^{-1}\mathbf{a} \right)\cdot\boldsymbol{\chi}} \label{eq:gaugesymmetrychi1}\\
  \implies\mathcal{T}^{-1}\tilde{\psi}^{\mathbf{a}}\mathcal{T}&=\mathcal{T}^{-1}\psi^{\mathbf{a}}e^{i\mathbf{a}\cdot\boldsymbol{\xi}}\mathcal{T}\nonumber\\ &=
  \left[ \psi^{\left( T^T\mathbf{a} \right)} \right]^\dagger e^{-i\pi\left( K^{-1}\mathbf{a} \right)\cdot\chi}e^{-i\mathbf{a}\cdot\boldsymbol{\xi}}.
  \label{eq:gaugesymmetrychi2}
\end{align}
We define the time reversal vector $\tilde{\boldsymbol{\chi}}$ in the new gauge in terms of the action of $\mathcal{T}$ on $\tilde{\psi^{\mathbf{a}}}$, analogous to equation \eqref{eq:gaugesymmetrychi1}.
\begin{align}
  \mathcal{T}^{-1}\tilde{\psi}^{\mathbf{a}}\mathcal{T} &=\left[ \tilde{\psi}^{\left( T^T\mathbf{a} \right)} \right]^\dagger e^{-i\pi\left( K^{-1}\mathbf{a} \right)\cdot\tilde{\boldsymbol{\chi}}}.
  \label{eq:timereversalnewgauge}
\end{align}
Combining equations \eqref{eq:gaugesymmetrychi2},\eqref{eq:timereversalnewgauge}, we get
\begin{equation}
  \tilde{\boldsymbol{\chi}}=\boldsymbol{\chi}+\frac{1}{\pi}K(1-T)\boldsymbol{\xi}\quad (\text{mod}\,\,2)
  \label{eq:gaugeequivalentchi}
\end{equation} 
Thus we would say $\boldsymbol{\chi}$ and $\tilde{\boldsymbol{\chi}}$ are gauge equivalent to each other.

The other constraint on $\boldsymbol{\chi}$ is determined by the action of $\mathcal{T}$ on local operators $\psi_{\text{local}}$ on the edge.
Local vertex operators are of the form, $\psi_{\text{local}}=e^{i\boldsymbol{\lambda}^TK\boldsymbol{\phi}}$; $\boldsymbol{\lambda}\in\mathbb{Z}^N$. Since the system under consideration is bosonic
\begin{align*}
  \mathcal{T}^{-2}\psi_{\text{local}}\mathcal{T}^2&=\psi_{\text{local}}.\nonumber\end{align*}
  However, we also know that \begin{align*}\mathcal{T}^{-2}\psi_{\text{local}}\mathcal{T}^2 &=\mathcal{T}^{-2}e^{i\boldsymbol{\lambda}^TK\boldsymbol{\phi}}\mathcal{T}^2=
  e^{i\boldsymbol{\lambda}^TK\boldsymbol{\phi}}e^{i\pi\boldsymbol{\lambda}^T\boldsymbol{\chi}}e^{-i\pi\left( T\boldsymbol{\lambda} \right)^T\cdot\boldsymbol{\chi}}.
\end{align*}
Since this must be true for all $\boldsymbol{\lambda}$, we find the constraint
\begin{equation}
  \left( \mathbf{I}-T^T \right)\boldsymbol{\chi}=0 \quad\text{mod }2.
  \label{eq:constraintchi}
\end{equation}
Equations \eqref{eq:gaugeequivalentchi},\eqref{eq:constraintchi} are very important in the definition of $\boldsymbol{\chi}$ and we will use them in the 
next section.
\subsection{Time reversal at the edge of bosonic fractional quantum spin Hall states}\label{subsec:chizero}
In the main text, we studied  the edge of a bosonic fractional quantum spin Hall system with $K$ matrix\\
$K^{\sigma\sigma'}_{2r\times 2r}= 
\begin{pmatrix}
 K_{r\times r}&0\\
 0&-K_{r\times r}
\end{pmatrix}
$, where
$\sigma=R,L=\uparrow,\downarrow$.\\
The charge vector for this system is
$\mathbf{t}^{\sigma}_{2r\times1}=
\begin{pmatrix}
	\mathbf{t}_{r\times 1}\\
	\mathbf{t}_{r\times 1}
\end{pmatrix}$.\newline
A suitable value of $T$  is
\begin{equation}
T^{\sigma\sigma'}_{IJ}=(\sigma_x)^{\sigma\sigma'}\delta_{IJ} =
\begin{pmatrix}
	0&\mathbf{1}_{r\times r}\\
	\mathbf{1}_{r\times r}&0
      \end{pmatrix}
\end{equation}
      It obeys $T^2=\mathbf{1}$; $T^{T}K^{\sigma\sigma'}T=-K^{\sigma\sigma'}$ and $T\mathbf{t^\sigma}=\mathbf{t^\sigma}$.\newline
Now, we need a time reversal vector
 $\boldsymbol{\chi^\sigma}=\begin{pmatrix}
	\boldsymbol{\chi}_{\uparrow}\\
	\boldsymbol{\chi}_{\downarrow}
\end{pmatrix}$ which obeys \eqref{eq:constraintchi}.\newline 
\begin{align}
   \left(\mathbf{I}-T^{T}\right)\boldsymbol{\chi^\sigma}=0\text{ mod }2
   &\implies\begin{pmatrix}
 1&-1\\
 -1&1
\end{pmatrix}\begin{pmatrix}
	\boldsymbol{\chi}_{\uparrow}\\
	\boldsymbol{\chi}_{\downarrow}
      \end{pmatrix}=0\text{ mod }2\nonumber\\
      \implies\,\,\boldsymbol{\chi}_{\uparrow}&=\boldsymbol{\chi}{\downarrow} \text{ mod }2
  \label{eq:chi1}
\end{align}
The vector $\boldsymbol{\chi}=
\begin{pmatrix}
  0\\
  0
\end{pmatrix}$ trivially satisfies this condition.\\
Next, we claim that {\emph{any}} valid time reversal vector $\boldsymbol{\chi}^{\sigma}$ (which satisfies equation \eqref{eq:chi1})
is gauge equivalent to $
\begin{pmatrix}
 0\\
  0
\end{pmatrix}$.\\ 
Using equation \eqref{eq:gaugeequivalentchi}
an equivalent statement is
\vskip -0.6 in
\begin{widetext}
\begin{align*}
  \exists \begin{pmatrix}
  \boldsymbol{\xi_{\uparrow}}\\
  \boldsymbol{\xi_{\downarrow}}
\end{pmatrix}\,\, :
  \begin{pmatrix}
  0\\
  0
\end{pmatrix}=\begin{pmatrix}
  \boldsymbol{\chi_{\uparrow}}\\
  \boldsymbol{\chi_{\downarrow}}
\end{pmatrix}+\frac{1}{\pi}\begin{pmatrix}
 K&0\\
 0&-K
\end{pmatrix}\begin{pmatrix}
 1&-1\\
 -1&1
\end{pmatrix}\begin{pmatrix}
  \boldsymbol{\xi_{\uparrow}}\\
  \boldsymbol{\xi_{\downarrow}}
\end{pmatrix}\quad\text{mod 2}\quad\Bigg|\quad\forall \begin{pmatrix}
  \boldsymbol{\chi_{\uparrow}}\\
  \boldsymbol{\chi_{\downarrow}}
\end{pmatrix}:\quad \boldsymbol{\chi}_{\uparrow}=\boldsymbol{\chi}_{\downarrow} \text{ mod }2
\end{align*}
\end{widetext}
This reduces to
\begin{equation*}
   \begin{pmatrix}
  \boldsymbol{\chi_{\uparrow}}\\
  \boldsymbol{\chi_{\downarrow}}
\end{pmatrix}=-\frac{1}{\pi} \begin{pmatrix}
  K\left( \boldsymbol{\xi}_{\uparrow}-\boldsymbol{\xi}_{\downarrow} \right)\\
  K\left( \boldsymbol{\xi}_{\uparrow}-\boldsymbol{\xi}_{\downarrow} \right)
\end{pmatrix} \quad\text{mod }2
\end{equation*}
A solution to this equation is some
$\begin{pmatrix}
  \boldsymbol{\xi_{\uparrow}}\\
  \boldsymbol{\xi_{\downarrow}}
\end{pmatrix}$ such that $\boldsymbol{\chi}_\uparrow=-\frac{1}{\pi}K(\boldsymbol{\xi}_\uparrow-\boldsymbol{\xi}_\downarrow)$. Since $\text{det}(K)\neq 0$ such a solution exists.
Furthermore, since $\boldsymbol{\chi}_\uparrow=\boldsymbol{\chi}_\downarrow$ mod 2, 
$\boldsymbol{\chi}_\downarrow=-\frac{1}{\pi}K(\boldsymbol{\xi}_\uparrow-
\boldsymbol{\xi}_\downarrow)$ mod 2 is automatically satisfied.\\
In conclusion we have shown that for the purposes of bosonic fractional quantum spin Hall states we can fix our gauge so that the time
reversal vector $\boldsymbol{\chi}^\sigma=0$, since all other choices of $\boldsymbol{\chi}^\sigma$ are gauge equivalent to it.
Thus from now on we can and henceforth we will use the gauge in which
\begin{equation}
  \mathcal{T}^{-1}\phi^{L/R}_{I}\mathcal{T}=\phi^{R/L}_{I}.
  \label{eq:trbosonfinal}
\end{equation}

\end{document}